\documentclass[times]{speauth}
\pdfoutput=1
\usepackage{dashrule}
\usepackage{amsthm}

\usepackage{hyperref}
\hypersetup{
  colorlinks=false  
}
\usepackage{float}
\usepackage{cancel}
\usepackage{algorithm}
\usepackage{multirow}

\usepackage[noend]{algorithmic} 
\newfloat{algorithm}{t}{lop}  

\usepackage{verbatim}
\def\ignore#1{}

\usepackage{colortbl}
\usepackage{siunitx}  
\usepackage{pgf}
\usepackage{tikz}
\usetikzlibrary{arrows,automata,decorations.markings,positioning,calc}
\usepackage{graphicx}
\usepackage{amssymb}
\usepackage{amsmath}

\usepackage{threeparttable}
\usepackage{url}
\usepackage[T1]{fontenc}
\usepackage{times}
\usepackage{verbatim}
\usepackage{subfig}
\usepackage{todonotes}

\usepackage{listings}
\usepackage{inconsolata}
\graphicspath{{gnuplot/}{.}}
\makeatletter
\def\input@path{{../Java/indexing/results/analysis/}{.}}
\makeatother

\newcommand{\thresh}{\vartheta}

\newcommand{\PGDVD}{PGDVD}

\newcommand{\IMDBthree}{IMDB-3gr}
\newcommand{\PGDVDtwo}{PGDVD-2gr}

\newcommand{\CensusIncome}{CensusIncome}
\newcommand{\Tweed}{TWEED}
\newcommand{\Weather}{Weather}

\newcommand{\scncnt}{\textsc{ScanCount}}
\newcommand{\looped}{\textsc{Looped}}

\newcommand{\mgopt}{\textsc{MgOpt}}
\newcommand{\dsk}{\textsc{DSk}}
\newcommand{\wheap}{\textsc{wHeap}}
\newcommand{\cdom}{\textsc{RBMrg}}

\newcommand{\wtwocti}{\textsc{w2CtI}}

\newcommand{\bstm}{\textsc{BSTM}}

\title{Compressed bitmap indexes: beyond unions and intersections}
\author{Owen Kaser\affil{1}\corrauth\ and Daniel Lemire\affil{2}}
\address{\centering \affilnum{1}Dept.\ of CSAS, 
University of New Brunswick,  Saint John, NB, Canada\\
\affilnum{2} LICEF, TELUQ, Universit\'e du Qu\'ebec, Montreal, QC, Canada}
\date{\today}
\corraddr{Owen Kaser, Dept.\ of CSAS, 
University of New Brunswick, 100 Tucker Park Rd, Saint John, NB E2L 4L5 Canada. email: owen@unbsj.ca}
\cgsn{NSERC}{99999}
\begin{document}

\runningheads{O. Kaser, D. Lemire}{Compressed bitmap indexes:  Beyond unions and intersections}

\cgsn{Natural Sciences and Engineering Research Council of Canada}{261437}

\begin{abstract}
Compressed bitmap indexes are used to speed up simple aggregate
queries in databases. Indeed, set operations like intersections, unions and
complements can be represented as logical operations (AND,OR,
NOT) that are ideally suited  for bitmaps. 
However, it is less obvious how to apply bitmaps
to more advanced queries. 
For example, we might seek
products in a store that meet some, but maybe not all, criteria. Such threshold queries generalize
intersections and unions; they are often used in 
information-retrieval and data-mining applications.
We introduce new
algorithms that are sometimes 
three
orders of magnitude faster than a na\"ive approach.
 Our work shows that
bitmap indexes are more
broadly applicable than
is commonly believed.
\end{abstract}


\maketitle

\section{Introduction}

There are many applications for bitmap indexes, from conventional
databases (e.g., Oracle~\cite{874730}) all the way to information
retrieval~\cite{Culpepper:2010:ESI:1877766.1877767} and
column stores~\cite{1083658}. They are  used
in data-warehouse platforms such as Apache Hive, LucidDB~\cite{thomsen2011survey}, Druid~\cite{druid2014} and Sybase~IQ~\cite{MacNicol:2004:SIM:1316689.1316798}. 

 We are primarily motivated by the
application of bitmap indexes to common databases (i.e., row
stores). In this case, it has long been established that bitmap
indexes can speed up several queries, e.g.,  joins~\cite{O'Neil:1995:MJT:211990.212001}, as well as
intersections and unions (e.g., \textsc{SELECT * WHERE A=1 AND
  B=2}). 
 
 Databases are commonly used for data mining and machine learning. An algorithm could seek to identify all movies that are ``similar'' to a target movie, or all customers that ``almost'' fit a given profile. Such queries
 need neither an intersection nor a union, but something in-between: a threshold function where only some of the criteria need to be satisfied. 
We aim to show that such queries (specifically \emph{Many-Criteria} queries and \emph{Similarity} queries, see \S~\ref{sec:advancedqueries}) can be answered efficiently using bitmap indexes.   
Because the result of the query is itself a bitmap, we can then further process it using the
standard operations permitted on bitmaps (OR, AND, XOR, NOT) to answer more complicated queries efficiently. 

Of course, the set of basic operations supported by bitmap indexes
 may be sufficient to synthesize any 
required function.  However, the efficiency of
such approaches is unknown.
 To our knowledge, the efficient computation of threshold functions over bitmaps
 has never been investigated in depth: the exception is Rinfret et al.~\cite{rinfret:bit-sliced-arithmetic} where two algorithms are compared on a related problem (top-$K$ queries). 

This paper considers several algorithms for threshold functions over
compressed bitmap indexes (see Table~\ref{tab:algos}).  Some of these algorithms are
novel (\looped{}, \wtwocti{}),  whereas other algorithms are adaptations of known algorithms that had
operated over sorted integer lists (\scncnt{}, \mgopt{}, \dsk{}) or over bitmaps (\bstm{}). 
(A companion report~\cite{symmetric-tr} considers additional 
algorithms that do not perform as well, and also considers the use of
uncompressed bitmaps.) 
The theoretical analyses of these alternatives, summarized in
Table~\ref{tab:complexity-rle-compressed}\footnote{The notation used
  throughout can be found in Table~\ref{tab:notation}.}, suggests that
there would be no single best algorithm for all cases, as the algorithms'
running times depend on different factors.
 Experiments described in
\S~\ref{sec:experiments} confirm this conclusion.  Thus one of our
contributions is a 
set of rules for automatically choosing algorithms.

Our work is organized as follows. In \S~\ref{sec:formulation}, we
formalize the problem. In \S~\ref{sec:relatedwork}, we present some
background material and related work. In \S~\ref{sec:advancedqueries},
we present the queries over database tables that we use for
benchmarking. In \S~\ref{sec:with-and-without-index}, we begin our experimental report by
showing that using a bitmap index, albeit na\"ively, is better than a
full table scan: the indexed version is anywhere from 1.1 to 6~times faster. In \S~\ref{sec:existingapproaches} and \S~\ref{sec:newapproaches}, we present our various
algorithms. Finally, in \S~\ref{sec:experiments} we assess them
experimentally and show that 
one can 
do significantly better
than a na\"ive approach:  up to
$1100\times$  
better, in one case. 
Over a large workload that we constructed, we could
more than triple
performance.  

\begin{table}
\caption{\label{tab:algos}Algorithms considered in this paper.}
\centering
\tabsize
\begin{tabular}{lllp{0.5\textwidth}} \toprule
Algorithm         & Source                                   & Section & Main idea\\ \midrule 
\scncnt           & \cite{Li:2008:EMF:1546682.1547171}       &  \S~\ref{sec:scancount} & Allocate array of counters, scan values while incrementing counters and, finally, scan array of counters for matching counts.\\
\mgopt            & \cite{Sarawagi:2004:ESJ:1007568.1007652,barbay2003deterministic} &  \S~\ref{sec:t-occurrence-algos} & Set aside the largest $T-1$~inputs, merge the remaining $N-T+1$~inputs using a heap, then look up matching values in the largest inputs.\\
\dsk           & \cite{Li:2008:EMF:1546682.1547171}       &  \S~\ref{sec:t-occurrence-algos} & Similar to \mgopt{}, but during the merger of the small inputs, some values are skipped; requires a tuning parameter.\\
\bstm           &  modified from \cite{rinfret:bit-sliced-arithmetic,253268}                              &  \S~\ref{sec:adding-circuit-algos} & Transforms the query into a Boolean circuit to be evaluated on the bitmaps.\\
\midrule 

\wtwocti          & novel                               &  \S~\ref{sec:mergeable-count-algos} & Merge inputs two-by-two starting with lowest-cardinality inputs while maintaining counters, prune results as early as possible.\\
\looped           &  novel                              &  \S~\ref{sec:looped-algo} & Allocates $T$~temporary bitmaps corresponding to the count values $1, 2, \ldots, T$, the first bitmap updates the first temporary bitmap, the second bitmap updates the first two temporary bitmaps, and so on. \\
\cdom             & novel, inspired by \cite{arxiv:0901.3751}     &  \S~\ref{sec:cdom-algo}& Using a heap, merge RLE-compressed words.\\
\bottomrule
\end{tabular}
\end{table}

\section{Formulation}\label{sec:formulation}

We take $N$  \emph{sorted} sets over a universe having $r$
distinct values.
For our purposes, we represent sets as bitmaps using $r$~bits. 
For example, if $N=2$ and $r=8$, we might have the sets $\{1,4,5\}$ and 
$\{4,5,7\}$ of integers in $[0,8)$ represented using the bitmaps $00110010$ and $10110000$,
where the least-significant bit represents the smallest value in the universe
(see \S~\ref{sec:bitmaps}).
 The  notation we use throughout is described in 
Table~\ref{tab:notation}.

\begin{figure}
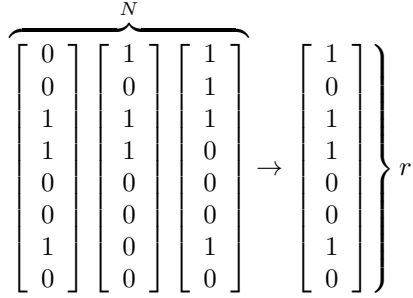
\centering

$\overbrace{  \left [ \begin{array}{c}
0\\ 
0\\
1\\
1\\
0\\
0\\
1\\
0\\
\end{array} \right ]
\left [\begin{array}{c}
1\\ 
0\\
1\\
1\\
0\\
0\\
0\\
0\\
\end{array}\right ]
\left [\begin{array}{c}
1\\ 
1\\
1\\
0\\
0\\
0\\
1\\
0\\
\end{array}\right ]}^{N}  \rightarrow \left . \left [\begin{array}{c}
1\\ 
0\\
1\\
1\\
0\\
0\\
1\\
0\\
\end{array}\right ] \right \} r$
\caption{\label{fig:basicexample}Solution to a threshold query with $T=2$ over $N=3$~bitmaps.}
\end{figure}

The sum of the cardinalities of the $N$ sets  is $B$: in our example $B=3+3=6$.
The cardinality of a set  is also given by the number of 1s in the corresponding bitmap. 
(By extension, the cardinality of a bitmap is the  number of 1s it contains.)
Therefore, the value $B$ is also the total number of 1s in all bitmaps.
 We apply a
threshold $T$ ($1\leq T \leq N$), seeking those elements that occur
in at least $T$~sets (see Fig.~\ref{fig:basicexample}). Because the cases $T=1$ and $T=N$ correspond to 
intersections and unions, which are well understood, we assume that $2 \leq T \leq N-1$.  
These queries are often called $T$-overlap~\cite{li2013fast, behm2009space}, 
$T$-occurrence~\cite{Li:2008:EMF:1546682.1547171,jia2012eti} or 
$T$-threshold~\cite{barbay2003deterministic,barbay2002adaptive} queries.

We can map a $T$-overlap query to a query over bitmaps using a Boolean
threshold function: given $N$~bits, the $T$-threshold function $\thresh(T, \{b_1, \ldots , b_N\})$ returns
true if at least $T$~bits are true; it returns false otherwise. For example, 
given $T=N$, such a function would just be a logical conjunction (AND) and given $T=1$, it would be a logical disjunction (OR).
That is, we have $\thresh(N, \{b_1, \ldots , b_N\}) = b_1 \land  \cdots \land b_N$ and  $\thresh(1, \{b_1, \ldots , b_N\}) = b_1 \lor  \cdots \lor b_N$.

A (unary) bitmap index over a table has as many bitmaps as there are distinct
attribute values (see Fig.~\ref{fig:examplebitmapindex}).  Each attribute value (say value $v$ of attribute $a$)
has a bitmap that encodes the set of row IDs that satisfy the predicate
 $a=v$.  A $T$-overlap query seeks row IDs that 
satisfy at least $T$ of $N$~predicates.
 Since each predicate is encoded as a bitmap, we need to compute
a bitwise threshold function over the $N$ chosen bitmaps.

\begin{figure}
\centering
{\small
\begin{tabular}{ll}
Name & City \\\hline
John & Montreal \\
Peter & Montreal \\
Jack & Toronto \\
Jack & Toronto \\
Jill & Toronto \\
Lucy & Paris \\
Mary & Toronto \\
\end{tabular}}$\rightarrow \overbrace{\left [ \begin{array}{c}
1\\ 
1\\
0\\
0\\
0\\
0\\
0\\
\end{array} \right ]}^{\text{Montreal}}
\overbrace{\left [ \begin{array}{c}
0\\ 
0\\
1\\
1\\
1\\
0\\
1\\
\end{array} \right ]}^{\text{Toronto}}
\overbrace{\left [ \begin{array}{c}
0\\ 
0\\
0\\
0\\
0\\
1\\
0\\
\end{array} \right ]}^{\text{Paris}}$
\caption{\label{fig:examplebitmapindex}Bitmap index of the attribute City.}
\end{figure}

Threshold functions are a subset of the \emph{symmetric Boolean functions}
(see \S~\ref{sec:boolean-functions-background}).
They include the majority function: given $N$~bits, the majority
function returns true when 
$1+\lfloor N/2\rfloor$
or more bits are true, and it returns
false otherwise. We can compute the majority function as any other threshold function. Other potentially useful generalizations include setting up a maximum (no more than $T$~bits are set) or setting a range (the number of set bits is in $[T_1,T_2]$). 
They can be rewritten in terms of threshold functions: e.g.,
we can determine whether at most $T$~bits are set by evaluating
$\thresh(N-T, \{\neg b_1, \ldots , \neg b_N\})$.
We do not consider such generalizations further. 

We denote the processor's native word length as
$W$ (typically\footnote{Common 64-bit PCs have SIMD instructions that work over 128-bit 
(SSE and AVX), 256-bit (AVX2) or 512-bit (AVX-512) vectors. These instructions 
might be used automatically by compilers and interpreters. Other general purpose processors in embedded or mobile
devices sometimes have a 32-bit word size. }
$W=64$). 
An uncompressed bitmap will
have $\lceil r/W \rceil $~words; given $N$~bitmaps, there are
 $N \lceil r/W \rceil $~words.  To simplify, we assume $\log N < W < r$ 
as well\footnote{In this paper, $\log n$ means $\log_2 n$.} 
as $\log r \leq W$, which
would typically be the case in the applications we envision.  
Also, we assume that $B \geq N$, 
which would be true if there is no bitmap containing only 0s: such \emph{empty} bitmaps could be virtually deleted without harm. 

\begin{table}
\caption{\label{tab:notation} Notation used in analyses.}
\centering
\tabsize
\begin{tabular}{ll} \toprule
Symbol & Meaning\\ \midrule
$A_i$ or $B_i$  & $i^{\mathrm{th}}$ bitmap\\
$|B_i|$& number of 1s in $i^{\mathrm{th}}$ bitmap\\
$B_i[j]$& value of the $j^{\mathrm{th}}$ bit in the $i^{\mathrm{th}}$ bitmap\\
$B$    & $\sum_i |B_i|$\\
$B'$   & number of 1s \textbf{not} in the $T-1$ largest bitmaps\\
\textsc{EWAHSize} &storage size in bytes of a collection of compressed bitmaps\\
$N$    & Number of bitmaps in the query\\
$r$    & length of bitmaps (largest index covered) \\
\textsc{RunCount} &number of runs of 0s and 1s in a collection of bitmaps\\
$T$    & Minimum threshold\\
$\thresh(T, \{b_1, \ldots , b_N\})$ & threshold function over bits $b_i$\\
$W$    & machine word size\\
$\oplus$, XOR & exclusive or \\
$\land$, AND & logical and \\
$\lor$, OR & logical or \\
$\neg$, NOT & logical negation \\ 
\bottomrule
\end{tabular}
\end{table}

\begin{table}[tbh]
\centering
\begin{threeparttable}[b]
  \caption{\label{tab:complexity-rle-compressed}Time and memory
    complexity of threshold algorithms over RLE-compressed bitmap
    indexes. 
    }
\tabsize
\begin{tabular}{lllp{3.5cm}} \toprule
Algorithm         &  Time complexity  & Memory       & Comment\\ \midrule
\scncnt           &  $O(r+B)$                           & $O(r)$                  & {\small Efficient access pattern }      \\
\mgopt            & $O(B' (\log(N-T)+T)+B-B')$          & $O(N)$                & Pruning can reduce $B$  \\
\dsk              & $O(B'(\log(N-T)+T)+B-B')$           & $O(N)$                & Pruning can reduce $B'$ \& $B$ \\
\bstm           & $O(Nr/W \times \log N)$               & $O(\log N \times r/W)$  & Note \tnote{1} \\
\midrule
\wtwocti          & $O(B(N-T))$                         & $O(B)$                  & \\
\looped           & $O(NTr/W)$                            & $O(Tr/W)$               & Note \tnote{2}       \\
\cdom             & $O(\textsc{RunCount} \times \log N)$& $O(N)$                  &      \\
\bottomrule
\end{tabular}
\begin{tablenotes}        
\item [1] $O(N\log N)$ basic bitmap operations are used, producing temporary results.  There are
$O(\log N)$ temporaries live at any time. 
An $O(Nr/W \times \log N)$ time bound ignores any benefits of compression for storage or processing.
\item [2] Fewer than $2NT$ basic bitmap operations are used, and $T$ temporary bitmaps
are used.  The bounds shown ignore compression's
benefits.
\end{tablenotes}
\end{threeparttable}
\end{table}

Our focus is on algorithms that run in main memory; we assume that the $N$
bitmaps involved in the threshold query have already been read into main memory.
Our memory bounds in Table~\ref{tab:complexity-rle-compressed} are based on
the \emph{additional} working memory required, not including the input and output.

\paragraph{A lower bound (and beating it):}

Towards a lower bound for the problem, note that if the output indicates
that $X$~entries meet the threshold, 
at least $TX$~1s 
have been observed
in the input.
If each such observation triggers $\Omega(1)$ work (as it does with
\scncnt\ (\S~\ref{sec:scancount}), when a counter is incremented),
this implies an $\Omega(TX)$ lower bound.  Barbay and
Kenyon~\cite{barbay2003deterministic} have established a
data-dependent lower bound for the problem, assuming the data is
presented in sorted arrays and using a model where comparisons are the
only allowed operations on array elements.
However, both bounds leave open
the possibility of using parallelism.  One such approach,
parallelization of \scncnt\ on GPUs, is described by Li et
al.~\cite{li2013fast}.  We can use bit-level parallelism (readily
available in bitmap inputs) to process several events per machine
operation. (See \S~\ref{sec:circuit-algos} and \S~\ref{sec:looped-algo}.)  
The bounds also leave open the
possibility of using Run Length Encoding (RLE), whereby many
consecutive events can be succinctly represented and processed
together.  Our compressed bitmap inputs are suitable for
such an approach: see \S~\ref{sec:cdom-algo}.

\section{Background and related work}
\label{sec:relatedwork}

We review some key concepts on compressed bitmaps, Boolean circuits,
and Boolean functions---especially symmetric and threshold functions.

\subsection{Bitmaps}
\label{sec:bitmaps}

We find bitmap indexes in several database systems, going as far back as the MODEL~204 database engine, commercialized  in 
1972~\cite{658338}. Most commonly, a bitmap index associates a bitmap (also called bitset or bit vector)  with every
attribute value $v$ of every attribute $a$;
the bitmap represents the predicate $a=v$. In the example of Fig.~\ref{fig:examplebitmapindex}, we see that we could identify all rows where the value of the attribute is either Montreal or Toronto by computing the bitwise OR between two bitmaps. Such bitwise operations can be computed quickly by most processors.
 In an experimental evaluation using the Oracle database system, Sharma found that a bitmap index is  preferable to a B-tree when the data is infrequently updated~\cite{oraclevivekbitmap}.

We consider compressed and uncompressed bitmaps.  The \emph{density} of a bitmap
is the fraction of its bits that are 1s.  A bitmap with low density is
\emph{sparse}, and such bitmaps arise in many applications. A bitmap with
density closer to 1 (perhaps \SI{5}{\percent} or more) is \emph{dense}.

\paragraph{Uncompressed Bitmaps}
An uncompressed bitmap represents a sorted set $S$ over $\{0,1,\ldots,r-1\}$ using  
$\lfloor (r+W-1)/W \rfloor$ 
consecutive words. 
The $W$ bits in the first word
record which values in $[0,W-1]$ are present in $S$.  The bits in the second word
record the values in $[W,2W-1]$ that are in $S$, and so forth.
 Within a word, the least-significant bit 
represents the smallest value.
For example, the set $\{1,2,7,9\}$ is represented as
10000110 00000010 with $W=8$. The first
word (10000110) represents the first 3~integers ($\{1,2,7\}$) whereas the second word (00000010) is used to store the value 9. The density is $\frac 4 {16}$.
The exact mapping between integers
and the bits within a word is unimportant, as long as it is always consistent 
(e.g., $\{1,2,7\}$ could be written as 10000110 or 01100001). 
The number of 1s is always equal to the cardinality of the set.

Uncompressed
bitmaps have the advantages of a fixed size (updates do not change the size)
and an efficient membership test.  However, if $r$ is large, the bitmap
occupies many words and uses much memory---even when 
representing a small set ($B \ll r$).

\paragraph{Compressed Bitmaps}

In a bitmap,  there are runs of consecutive 0s and  runs of consecutive 1s.
The number of such runs is called the \textsc{RunCount} of a bitmap, or
of a collection of bitmaps~\cite{rlewithsorting}.  For example, in the bitmap index illustrated by Fig.~\ref{fig:examplebitmapindex}, there are $2+4+3=9$~runs. In the unary bitmap index
of an attribute containing $N$~distinct attribute values, given that there are $r$~rows, the number
of runs must be between $3N-2$ and $2r+N-2$. Correspondingly, for $r \gg N$, 
the average length of
the runs is between $\approx r/3$ and $\approx N/2$.
In many situations where bitmaps are generated, we expect to find many long runs (e.g., with length greater than $W$).
  
Though there are alternatives~\cite {navarro2012fast}, the most
popular compression techniques are based on the (word-aligned) RLE
compression model inherited from Oracle (BBC~\cite{874730}):
WAH~\cite{wu2008breaking},
Concise~\cite{Colantonio:2010:CCN:1824821.1824857},
PLWAH~\cite{Deliege:2010:PLW:1739041.1739071},
EWAH~\cite{arxiv:0901.3751}, COMPAX~\cite{netfli},
VAL-WAH~\cite{guzuntunable}, among others. 
The $r$~bits of the bitmap are partitioned into sequences of $W'$ consecutive bits,
where $W' \approx W$ depends on the technique used; for
EWAH, $W' = W$; for WAH, $W'=W-1$.
When such a sequence contains only 1s or only 0s, 
it is a \emph{fill}~word,
otherwise 
it is a  \emph{dirty} word.
For example, using $W'=8$ , the uncompressed bitmap $00000000 01010000$ contains two words,
a fill word ($00000000$) and a dirty word ($01010000$).
Techniques such as BBC, WAH or EWAH typically use special
marker words to compress long sequences of identical 
{fill}~words.
When accessing these formats, it may be necessary to read
every compressed word to determine whether it indicates a sequence of fill words,
or a 
{dirty}~word. 
The EWAH format~\cite{arxiv:0901.3751} supports a limited form of skipping because
it uses marker words not only to mark the length of the sequences of
fill words, but it also uses these markers to indicate the lengths of
the sequences of consecutive dirty words. Because of this feature, 
one can 
skip 
sequences  
of dirty words when using EWAH\@.

Though there are many good compressed formats to choose from, we have
picked EWAH\@. In a benchmark between various formats where the authors
used our implementation (the JavaEWAH library~\cite{JavaEWAH}), Guzun et
al.~\cite{guzuntunable} found that ``Although EWAH does not compress
well, (\ldots) it offers the best query time for all 
distributions.''   Moreover,
EWAH is used in a major data  database system (Apache Hive).
We refer the reader to previous work for the exact format specification~\cite{arxiv:0901.3751}.

Compressed bitmaps are often appropriate for storing sets that cannot be
efficiently handled by uncompressed bitmaps.  For instance, consider
the bitmap consisting of a million 0s followed by a million 1s.
This data has two runs ($\textsc{RunCount}=2$) but a million 1s.
It can be stored using EWAH in only a few words.

However, some RLE compressed bitmaps are not efficient for storing
extremely sparse data that does not have dense clusters.
For instance, consider EWAH: sparse data with very long runs of 0s 
between elements will result
in a marker word and a dirty word for each 1 bit. Because EWAH uses
64-bit words by default, we would use 128~bits per element. This would be 
less efficient than explicitly listing the set elements (e.g., 32~bits) by 
a factor of 4. Observe, however, that using 
(compressed) bitmaps for such sets is likely inefficient in any case:
bitmaps are efficient due to bit-level parallelism when there are many
words containing a mix of 1s and 0s.

Software libraries for compressed bitmaps will typically include
an assortment of basic Boolean operations that operate directly
on the compressed bitmaps.  One would expect to find operations
for AND, OR, and often one finds XOR, ANDNOT, and NOT\@.  
EWAH, like most other RLE-based formats, allows the operations
AND, OR, XOR and ANDNOT
between two compressed bitmaps ($B_1$ and $B_2$) 
to execute in time $O(\textsc{EWAHSize}(B_1)+\textsc{EWAHSize}(B_2))$.
Moreover, the output of such an aggregate has compressed size bounded by
the size of the input ($\textsc{EWAHSize}(B_1)+\textsc{EWAHSize}(B_2)$).
(For AND, the output is bounded by 
$\min(\textsc{EWAHSize}(B_1),\textsc{EWAHSize}(B_2))$.)

Some libraries
support only binary operations, whereas others support \emph{wide}
queries: for instance, a wide AND would allow us to intersect
four bitmaps in a single operation, rather than having to AND
bitmaps together pairwise.   Explicit support for wide operations can 
allow for better performance~\cite{1316694}. Threshold functions are wide 
queries when $N>2$. 

Our complexity analysis (Table~\ref{tab:complexity-rle-compressed}) assumes
that we can iterate over  the 1s in a compressed bitmap in
$\Theta(1)$ time each.
We can indeed iterate over the 1s in a compressed EWAH bitmap quickly.
 Runs of fill words are not problematic: 
e.g., 64-bit EWAH uses 32-bit counters for the length of such runs, so runs of 
up to $2^{32} \times 2^W$ identical bits can be marked with a single marker word.
Moreover, we can also extract 1s from dirty words quickly. In Java, we can use  the
\texttt{Long.numberOfTrailingZeros} function and a simple loop: this function 
is commonly compiled to efficient machine instructions by the JVM (e.g., \texttt{bsr} on Intel and AMD processors).

\subsection{Boolean Functions  and Circuits}
\label{sec:boolean-functions-background}

A Boolean function is a function of the form $f: \{0,1\}^k \to \{0,1\}$.
For relevant background on Boolean functions, see Knuth~\cite{KnuthV4A}. 
A Boolean circuit over some basis (e.g., AND, OR, NOT) is a  directed acyclic graph
where each vertex is either a  basis function or an input, and where some of the vertices are outputs.
Boolean functions can be computed by Boolean circuits.
As discussed in \S~\ref{sec:formulation},
some Boolean functions  are \emph{symmetric}.
These functions are unchanged under any permutation
of their inputs.  I.e., a symmetric function is completely determined
if one knows the number of 1s (the Hamming weight) in its
inputs.  An example symmetric function outputs 0 $\iff$ the
Hamming weight of its inputs is a multiple of 2: this is the XOR
function.

\subsection{Threshold Functions}

Threshold functions, in the guise of $T$-overlap queries,
 have been used for approximate searching.
Specifically,  Sarawagi and Kirpal~\cite{Sarawagi:2004:ESJ:1007568.1007652}
show how to avoid unnecessary and expensive pairwise distance computations (such
as edit-distance computations) by using threshold functions to screen out
items that cannot be approximate matches.  Their observation was that 
strings $s_1$ and $s_2$ must have many ($T$)  $q$-grams in common, if they have
a chance of being approximate matches to one another.  Given $s_1$ and
seeking suitable $s_2$ values, 
we take the set of $q$-grams of $s_1$.  Each $q$-gram is associated with
a set of the words (more specifically, with their row IDs) that
contain that $q$-gram at least once.  Taking these $N$ sets, we
use  a threshold function to determine values $s_2$ that can be compared
more carefully against $s_1$.  Using $q$-grams, Sarawagi and Kirpal showed that
$T=|s_1| + q - 1 - k q$ will not discard any string that might be within
edit distance $k$ of $s_1$. 
 In applications where $k$ and $q$ are small
but the strings are long, this will create queries where $T \approx N$.
(Similar formulae are known for Jaccard, cosine and dice
similarities~\cite{Li:2008:EMF:1546682.1547171,Sarawagi:2004:ESJ:1007568.1007652}.)

Closely related to $T$-overlap queries, 
we have Opt-threshold queries~\cite{barbay2003deterministic,tellez2013succinct}.
In these queries, $T$ is unspecified: the algorithm is responsible for
choosing the largest threshold value that
leads to a non-empty result. We could further generalize such queries by asking for the
largest  value $T$ such that the result of the $T$-overlap query 
contains at least $K$~elements.  ``Top-$K$'' versions of the 
problem~\cite{rinfret:bit-sliced-arithmetic} are closely related, but
are not symmetric bitwise Boolean 
operations---if the Opt-threshold result yields two elements,  a top-1
query will return only one of them, despite both meeting the same
threshold.  

\section{Advanced queries}\label{sec:advancedqueries}

To obtain results
that correspond to a practical applications of bitmap indexes, we focus on using threshold functions
over bitmap indexes to answer 
 two different types of queries, \emph{Many-Criteria} queries
and \emph{Similarity} queries.

\paragraph{Many-Criteria Queries:} The first type of query has a set
of criteria, and we are seeking those records that meet some minimum
number of the criteria, but perhaps not all. E.g., 
consider a
query that might be typical of some human-resources system (in
pseudo-SQL).  

\lstset{language=SQL}
\begin{lstlisting}
SELECT * FROM table WHERE Gender="F" AND 
  (City="Montreal" OR City="Vancouver") AND 
  experience>=24 AND education>=college;
\end{lstlisting} 
If it corresponds to an application where we filter job
candidates, maybe applying all constraints at once could lead to a
small (or empty) result set. Or maybe we want to include exceptional candidates
who fail to satisfy a few conditions.   So we are willing to relax the
constraint somewhat, by maybe requiring that only three of the constraints 
hold, as in the following example.

 \begin{lstlisting}
SELECT * FROM table WHERE  
    CASE WHEN Gender="F" THEN 1 ELSE 0 END
  + CASE WHEN City="Montreal" THEN 1 ELSE 0 END
  + CASE WHEN City="Vancouver" THEN 1 ELSE 0 END
  + CASE WHEN experience>=2 THEN 1 ELSE 0 END
  + CASE WHEN education >= college THEN 1 ELSE 0 END
    >= 3;
\end{lstlisting}

\paragraph{Similarity Queries:}
The second type of query presents a prototypical item.  We determine
the criteria that this item meets, and then seek all items that meet
(at least) $T$ of these criteria.  For example, if a user liked a
given movie, he might be interested in other similar movies (e.g.,
same director, or same studio, or same leading star, or same date of
release). As part of a recommender system, we might be interested in
identifying quickly all movies satisfying at least $T$ of these
criteria.  This might be viewed as setting a threshold on the Hamming
distance between tuples. 

Once the
criteria have been defined, SQL can
handle 
the rest of the query, as in the 
previous example.
Critchley~\cite{crit:similaritySQL} proposes an alternative SQL-only solution using
joins and SQL aggregation.
We consider the evaluation of such external-memory approaches outside our current scope.
Similarity queries have been used with approximate string
matching~\cite{Li:2008:EMF:1546682.1547171,Sarawagi:2004:ESJ:1007568.1007652}.
In this case, items are small chunks of text, and the occurrence of a
particular 3-gram (a sequence of 3~consecutive letters) is a criterion.
In that previous work, an index maps each
$3$-gram to a sorted list of integers that specify the chunks of text
containing it.  More recent work by others~\cite{ferr:duplicates-qgrams,
montanari2012near} solves similar problems using bitmaps, one for
each $2$-gram.

A generalization of a Similarity query presents \emph{several} prototypical items,
then determines the criteria met by at least one of them.   We then proceed
as before, finding all items in the database that meet at least $T$ of the
criteria.   If there are $n$~prototypes, we have a ``Similarity($n$)'' query.

 Assuming one has a bitmap
index, can one answer Many-Criteria  and Similarity queries
better than using the row-scan that would be done by a
typical database engine? One of our contributions is to show that it is indeed
the case. In \S~\ref{sec:with-and-without-index}, we show that a simple bitmap-based   
algorithm (\scncnt{}, see \S~\ref{sec:scancount})
 is able to outperform a row scan (e.g., by a factor of 6).  Then in \S~\ref{sec:experiments}
we show that other bitmap-based algorithms can outperform this simple  approach (\scncnt),
sometimes by hundreds of times.

\subsection{An index is better than no index}
\label{sec:with-and-without-index}

Could a simple T-occurrence query can be more effectively answered without
using a bitmap index? Before continuing our investigation with various
novel algorithms, we want to establish that bitmap indexes can accelerate 
some T-occurrence queries. Our purpose is merely motivational: detailed
experiments, including a description of our queries and datasets is given in \S~\ref{sec:experiments}.

\begin{algorithm}
\centering
\begin{algorithmic}[1]
\REQUIRE A table with $D$ attributes.  A set $\kappa$ of $N \leq D$ attributes, and for each such attribute a desired value. Some threshold $T$.
\STATE Create an initially empty set $s$ 
\FOR{each row in the table}
\STATE{counter $c \leftarrow 0$}
\FOR{for each attribute $k$ in $\kappa$}
\IF{attribute $k$ of the row has the desired value}
\STATE increment $c$
\ENDIF
\ENDFOR
\IF{$c\geq T$}
\STATE add the row (via a reference to it) to $s$
\ENDIF
\ENDFOR
\STATE return the set of matching rows, $s$
\end{algorithmic}
\caption{\label{alg:rowstore} Row-scanning approach over a row store.
}
\end{algorithm}

As a reference, we use a full table scan (see Algorithm~\ref{alg:rowstore}),
where the table is stored in RAM\@. 
To test the basic usefulness of a bitmap index, we use a simple
algorithm (\scncnt{}, see \S~\ref{sec:scancount} for details): 
we create an array of $r$~counters initialized to zero. Then
the bits
of each bitmap are scanned in sequence, one bitmap at a time. When
a 1-bit is found, the corresponding counter is incremented. The algorithm
concludes with a full scan of the all counters.

We made 30 trials,
on each of the datasets \CensusIncome, \Weather\ and \Tweed.  These
are described in \S~\ref{sec:real-datasets}
and have 42, 19 and 53 attributes, respectively.
We randomly chose one value per attribute and randomly chose a threshold 
between 1 and the number of attributes, exclusively.
This query corresponds to a Many-Criteria query.   
Table~\ref{tab:with-and-without-index} shows that  using 
an EWAH index for this query was 4--6 times
 faster than scanning the table. The advantage persisted, but was smaller, when we did a
Similarity query against a randomly chosen row.
It is reassuring
that a bitmap index using \scncnt{}  answered our queries faster  than
they would be computed from the base table. It remains to determine whether we can 
surpass \scncnt{}.
Section~\ref{sec:experiments} shows that two 
algorithms can run at least $1000\times$ 
faster than \scncnt\ on certain queries, although
speedups of $3\times$  to $5\times$ seem  more typical.

\begin{table}
\caption{\label{tab:with-and-without-index} Total time (ms) 
required for queries in our workload.\\  
Top: Many-Criteria query.  Bottom: Similarity query. 
}
\centering
\tabsize

\begin{tabular}{cSSS} \toprule
         & \multicolumn{1}{l}{\CensusIncome}   &  \multicolumn{1}{l}{\Weather}  & \multicolumn{1}{l}{\Tweed}\\ \midrule
EWAH \scncnt        &  109              & 201          & 6 \\
Row Scan (no index) &  487              & 1212         & 23 \\
Row Scan/\scncnt  (\%)  &  450              & 600          & 380 \\
 & & & \\
EWAH \scncnt        &  327              & 508          & 20\\
Row Scan (no index) &  557              & 1344         & 22\\
Row Scan/\scncnt  (\%)  &  170              & 260          & 110 \\
\bottomrule
\end{tabular}
\end{table}

\section{Existing approaches for Threshold Functions}
\label{sec:existingapproaches}
We next present several different approaches to computing threshold functions
that have been proposed in the literature.  Several
generalize  to handle all symmetric functions, and several can be modified
to solve Opt-threshold queries.

\subsection{Counter-based approaches} \label{sec:scancount} 

In information retrieval, it is common practice to solve threshold queries using sets of counters~\cite{Perry01021983}.
The simple \scncnt\ algorithm of Li et al.~\cite{Li:2008:EMF:1546682.1547171}
(previewed in \S~\ref{sec:with-and-without-index}) 
uses an array of counters, one counter per item. 
 The input is scanned, one bitmap
at a time. If an item (as a bit set to 1) is seen in the current bitmap, its counter is incremented. 
 A final pass
over the counters can determine which items have been seen at least $T$~times.  In our case, items correspond to positions in the bitmap.
If the maximum bit position is known in advance, if this position is not
too large,  and if one can efficiently iterate over
 the bit positions in a bitmap, 
then \scncnt\ is easily implemented.  These conditions are frequently met when the bitmaps
represent the sets of row IDs in a table that is not exceptionally large.

\scncnt{} is part of a family of  counter-based approaches that have the
characteristic that they count the 
occurrences of each item.
They can handle arbitrary symmetric functions, since one
can provide a user-defined function mapping $[0,N]$ to Booleans.  However,
some counter-based approaches can be optimized specifically to 
compute threshold functions (see \S~\ref{sec:mergeable-count-algos}).

To analyze \scncnt , note that it uses  $\Theta(r)$ counters.  We assume $N < 2^W$, so
each counter occupies a single machine word.  Even if counter initialization can be 
avoided (see Li et al.\ for details)
the algorithm compares each counter 
against $T$.  
Also, the total number of counter increments is $B$.  Together, these imply a
time complexity of $\Theta(r+B)$ and a space complexity of $\Theta(r)$.
Aside from the effect of
$N$ on $B$ (on average, a linear effect), note that this algorithm does not depend on~$N$.
(Li et al.~\cite{Li:2008:EMF:1546682.1547171} also present an alternative \scncnt\ algorithm that generates an unsorted list
in $O(B)$~time.
Generating a RLE-compressed bitmap would require sorting this output, and
this could be a major overhead for queries with large outputs.  Thus we do not consider this variation.)

The \scncnt\ approach fits modern hardware well:
 the counters are accessed in sequence, during the $N$
passes over them when they are incremented.  
Experimentally, we found that using 8-bit \texttt{byte} counters when $N < 128$
usually brought a small (perhaps \SI{15}{\percent})
 speed gain compared with 32-bit \texttt{int} counters.
Perhaps more importantly, 
this also quarters the memory consumption of the algorithm.
One can also experiment with other memory-reduction techniques:
e.g.,
if $T < 128$, 
one could use a saturating 8-bit counter.
Experimentally, we found that the gains usually were less than the losses
that come from the additional conditional check required to ensure saturation.
Based on our experimental results,
the \scncnt\ 
implementation used in \S~\ref{sec:experiments} switches
between \texttt{byte}, \texttt{short} and \texttt{int} counters based on
$N$, but does not use the saturating-count approach.

\scncnt\ fails when the bitmaps have extreme $r$ values.
If we restrict ourselves to bitmaps that arise within a bitmap index, this implies
that we have indexed a table with an extreme number of rows.
However, instead of using $r$~counters, we could use a small number and effectively partition the problem: choose a fixed number of counters $r'$ and execute \scncnt{} $\lceil r/r' \rceil$~times, always reusing the same counters. We exploit this idea in \S~\ref{sec:cdom-algo} with the  \cdom{} scheme.

It is easy to obtain an Opt-Threshold algorithm: \scncnt\ begins
as usual and obtains the
$r$ counters.  $T$ is the maximum value in the counters, and the algorithm then returns
those elements whose counters equal $T$. 

\subsection{T-occurrence algorithms for integer sets}
\label{sec:t-occurrence-algos}

Prior work~\cite{Li:2008:EMF:1546682.1547171,Sarawagi:2004:ESJ:1007568.1007652}
has studied the case when the  data is presented as sorted lists of integers
rather than bitmaps.  
 We  consider the following T-occurrence 
algorithms:  \wheap~\cite{Sarawagi:2004:ESJ:1007568.1007652},
\mgopt~\cite{Sarawagi:2004:ESJ:1007568.1007652,barbay2003deterministic},
and \dsk~\cite{Li:2008:EMF:1546682.1547171}.
For full details of these algorithms, see the papers that introduced them.
All can be viewed as modifications to the basic \wheap\ approach.
This approach essentially uses an $N$-element min-heap that contains
one element per input. 
Using the heap, it merges the
sorted input sequences.  As items are removed from the heap, we count duplicates
and thereby know which elements had at least $T$~duplicates.
This approach can be generalized to compute any symmetric function, but it
requires that we process the 1s in each list, inserting
(and then removing) the position of each into an $N$ element min-heap.  
The total time cost is thus $O(B \log N)$ for sorted lists.

The \wheap\ approach has been shown to have worse performance than
\mgopt\ or \dsk~\cite{symmetric-tr,Li:2008:EMF:1546682.1547171,Sarawagi:2004:ESJ:1007568.1007652} and thus is not considered further.

The remaining algorithms are also based around heaps (\mgopt{} and \dsk{}), but they are designed
to exploit characteristics of real data, such as skew, that allow us
to skip certain input elements.  In contrast with other algorithms (e.g., \wheap{}, \cdom{} and
\scncnt{}),
\mgopt{} and \dsk{} do not generalize to arbitrary symmetric functions because such
functions
preclude skipping any input. 
This is illustrated by the (wide) XOR function, whose output always
depends on all input bits---knowing all but one input bit is never
enough to determine the output.

\paragraph{Algorithm \mgopt :}
Sarawagi and Kirpal's \mgopt\ algorithm~\cite{Sarawagi:2004:ESJ:1007568.1007652} sets aside the largest \mbox{$T-1$~inputs}. 
Any item contained
only in these inputs cannot meet the threshold.  Then it uses an
approach similar to \wheap\ with threshold 1 on the smallest $N-T+1$~inputs. 
For each item found in the smallest inputs,
say with count $t$, the algorithm checks whether at least $T-t$~instances of
the item are found in the largest $T-1$~inputs.  The items are checked 
in the largest inputs in ascending sequence.  If one of the largest inputs is checked
for occurrence of item $x$, and the next check is for the occurrence of
item $y$, we know that $y>x$. Items between $x$ and $y$ in
the big input will never be needed, and can be skipped over without
inspection.  
Whereas we use bitmaps as inputs, Sarawagi and Kirpal use 
sorted lists of integers as inputs.  Thus they can use 
a doubling/bootstrapping binary search to find the smallest
value at least as big as $y$, without needing to scan all values between
$x$ and $y$. The portions skipped have been pruned.

As noted in \S~\ref{sec:bitmaps}, providing random access is not a
standard part of a RLE-based compressed bitmap library, although it is
essentially free for uncompressed bitmaps.  However, with certain
compressed bitmap indexes 
one can
``fast forward'', skipping
portions of the index in a limited way: the JavaEWAH library~\cite{JavaEWAH} uses the fact that we
can skip runs of dirty words (e.g., when computing intersections). 

To bound the running time, we can distinguish the $B-B'$ 1s in the
$T-1$ largest bitmaps from the $B'$ 1s in the remaining $N-T+1$ bitmaps.
A heap of size $O(N-T+1)$ is made of the $N-T+1$ remaining bitmaps, and 
$O(B')$ items will pass through the heap, at a cost of 
$O(\log(N-T+1))$ each. As each item is removed from the heap,
it will be sought in $O(T)$ bitmaps. 
Because the items sought
are in ascending order, the $T-1$ bitmaps will each be processed
in a single ascending scan that handles all the searches.
Each of the $B-B'$ 1s in the remaining bitmaps should cost us
$O(1)$ effort.  
Thus we obtain a bound of $O(B' (\log(N-T+1)+T)+B-B') =
O(B' (\log(N-T)+T)+B-B')$ for 
the time complexity of \mgopt.

A 
similar
algorithm was earlier presented by Barbay and Kenyon~\cite{barbay2003deterministic}.
Any input may appear in their heap, but at any time there will be
$T-1$ inputs that are not in the heap.  
Setting aside
the \emph{largest} items (as with Sarawagi and Kirpal) seems like a useful
enhancement. Indeed, consider our complexity bound of  $O(B' (\log(N-T)+T)+B-B')$:
each of the $B'$ elements has a multiplicative 
cost factor of $\log(N-T)+T$ 
 whereas each of the other
$B-B'$~elements has a cost factor of $1$. 
This reflects the fact that the $B'$~elements are stored in a heap
whereas the $B-B'$~elements are merely accessed sequentially.
 Thus we prefer to minimize
$B'$, which is done by setting aside the largest bitmaps. 

Our analysis does not take fully into account the effect of pruning,
because 
we might be able to skip many
of the $B-B'$ 1s as we search forward through the largest $T-1$
bitmaps.
Since these are the \emph{largest} bitmaps, if $T$ is close
to $N$ or if the sizes (number of 1s) in the bitmaps vary
widely,  pruning could make a large difference.  This depends
on the data.  Barbay and Kenyon
present a detailed running-time analysis (with input as 
sorted integer lists) in terms of a ``$t$-alternation'' 
parameter for the problem instance.  It matches their comparison-based
lower bound for the problem in many cases, and in all cases it
is within a factor of $O(\log(N-T+1))$ of the optimal complexity.

Barbay and Kenyon also describe how to obtain an Opt-threshold
algorithm from any $T$-overlap algorithm by successively trying $T=N$, $T=N-1$, \ldots
until a non-empty answer is obtained.  Although na\"ive,  the
empty $T$-overlap queries have a predictable cost 
for \mgopt\ (no worse than the final query),
 whereas a binary search for $T$
may make some more expensive queries. 

\paragraph{Algorithm \dsk :} 
Algorithm \dsk\ is essentially a hybrid of \mgopt\ and another pruning
algorithm called \textsc{MergeSkip.} 
\textsc{MergeSkip}~\cite{Li:2008:EMF:1546682.1547171} is like \wheap{} except that, 
when removing copies of an item from the heap, if there are not 
enough copies to meet the threshold, we remove  
some extra items.  This is done in such a way that the extra items removed
(and not subsequently re-inserted) could not possibly meet the threshold.
 (\textsc{MergeSkip} is not described further here, because its
performance is worse than \dsk{}~\cite{symmetric-tr,Li:2008:EMF:1546682.1547171}).  Algorithm \dsk{} processes the heap
as in \textsc{MergeSkip}, while it sets apart the  largest bitmaps as in \mgopt{}.
However, rather than following \mgopt\ and always
setting apart the $T-1$ largest sets, it chooses the $L$ largest sets
where $L$ is a tuning parameter.
Li et al.\ 
determine
another tuning parameter $\mu$  experimentally,
for a workload of queries against a given
dataset.  From $\mu$ and the length of the longest input, 
Li et al.\ use a heuristic formula
 for $L$ (see \S~\ref{sec:pickmu}). 
With a suitable $L$, we would not expect \dsk\ to perform
significantly worse than \mgopt. 

Our running-time complexity bound for \dsk\ is identical to that for \mgopt, and based
on the same reasoning that ignores pruning.
We cannot easily account for the pruning
opportunities that \dsk\  inherits from \textsc{MergeSkip} and \mgopt.  
However, as with \mgopt, data-dependent 
pruning could reduce the $B-B'$ term.  As with \textsc{MergeSkip}, the multiplicative
$B'$ factor can be reduced by data-dependent pruning~\cite{symmetric-tr}.

Considering memory, note that \mgopt\ and \dsk\ partition the inputs into
two groups.  Regardless of group, each compressed bitmap input will have an iterator constructed
for it.  
The first group also go into a heap that accepts one element per input.
Thus we end up with a memory bound of $O(N)$.

\subsection{Boolean synthesis} 
\label{sec:circuit-algos}

 A typical bitmap implementation provides a set of
basic operations, typically AND, OR, NOT, XOR and
sometimes ANDNOT\footnote{The 
x86 extensions SSE2 and AVX2 support AND NOT, as do several bitmap libraries (EWAH included).
Specifically, Intel has the \texttt{pandn} and \texttt{vpandn} instructions; 
however it does not appear
the standard x86 instruction set has a corresponding instruction.
}.
Since
one can 
synthesize any Boolean function using AND, OR and NOT operations in
combination,  any desired 
bitwise  
bitmap function can be ``compiled'' into a sequence
of primitive operations.
For instance, the threshold functions over individual bits are, for $N=3$,
\begin{itemize}
\item ($T=1$)\, $\thresh(1, \{b_1, b_2, b_3\})= b_1 \lor b_2 \lor b_3$, 
\item ($T=2$)\, $\thresh(2, \{b_1, b_2, b_3\})= (b_1 \land  b_2) \lor (b_2 \land  b_3) \lor (b_1 \land  b_3)$,
\item ($T=3$)\, $\thresh(2, \{b_1, b_2, b_3\})=  b_1 \land b_2 \land b_3$.
\end{itemize}
As a bitwise operation over EWAH bitmaps \texttt{B1}, \texttt{B2}
and \texttt{B3}, we have the corresponding Java expressions:
\begin{itemize}
\item ($T=1$)\, \verb+EWAHCompressedBitmap.or(B1,B2,B3)+, 
\item ($T=2$)\, \verb+EWAHCompressedBitmap.or(B1.and(B2), B2.and(B3), B1.and(B3))+,
\item ($T=3$)\, \verb+EWAHCompressedBitmap.and(B1,B2,B3)+.
\end{itemize}
Of course, these are examples: several Boolean expressions are equivalent to a given threshold function, and some are more efficient than others.
For example, we can also write  $\thresh(2, \{b_1, b_2, b_3\})= (b_2 \land (b_1  \lor   b_3)) \lor (b_1 \land  b_3)$---saving one Boolean operation over the alternative ($(b_1 \land  b_2) \lor (b_2 \land  b_3) \lor (b_1 \land  b_3)$).

In \S~\ref{sec:adding-circuit-algos}~and~\ref{sec:looped-algo} we
introduce threshold algorithms \bstm\ and \looped\ that 
 synthesize the desired bitmap function
from standard bitmap operations (binary AND, OR, XOR, and ANDNOT operations).
One major advantage is that this
approach allows us to use a bitmap library as a black box, although it is
crucial that the primitive operations have efficient algorithms and
implementations. O'Neill and Quass~\cite{253268} and
Rinfret et al.~\cite{rinfret:bit-sliced-arithmetic} implicitly used
this idea when doing arithmetic and comparison
operations bitwise over a bit-sliced index.  They note the opportunities for
bit-level parallelism that arise.
For example, the expression $\thresh(2, \{b_1, b_2, b_3\})= (b_1 \land  b_2) \lor (b_2 \land  b_3) \lor (b_1 \land  b_3)$ can actually compute 64~thresholds using only 5~bitwise operations  on a 64-bit architecture.
Without bit-level parallelism, we would need at least $3\times 64 / 2=96$~binary 
operations so that 
each input is used once:  the benefits of bit-level parallelism are at least a factor of $96/5=19.2$ in this case.

Unfortunately, it is computationally infeasible
to determine the fewest required primitive operations
that realize a desired Boolean function,
except in the simplest cases~\cite{KnuthV4A}. In any case, for RLE compressed bitmaps, 
the relative costs of the primitive operations depend on the data.

\subsubsection{Adding: The \bstm\ algorithm.}
\label{sec:adding-circuit-algos}

Rinfret et al.~\cite{rinfret:bit-sliced-arithmetic} used the Boolean synthesis
approach to solve a problem closely related to thresholding.  
In their information-retrieval problem, one seeks the top $k$ documents that best match
a set of keywords.  The input is provided as a collection of bitmaps, one
bitmap for each keyword. Set bits in a  bitmap indicate the presence of 
the keyword in a document.  The result of the query is a bitmap with $k$
bits set.

While the ``top-$k$'' aspect means that the required computation is not
a bitwise Boolean function, their method of solution can be adapted to solve
our threshold problem, leading to the following algorithm, \bstm.

\begin{figure}
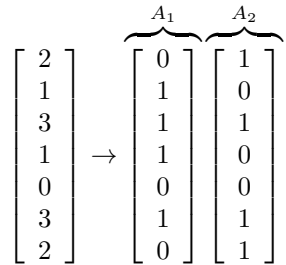

\centering
$\left [ \begin{array}{c}
2\\ 
1\\
3\\
1\\
0\\
3\\
2\\
\end{array} \right ]
\rightarrow 
\overbrace{\left [ \begin{array}{c}
0\\ 
1\\
1\\
1\\
0\\
1\\
0\\
\end{array} \right ]}^{A_1}
\overbrace{\left [ \begin{array}{c}
1\\ 
0\\
1\\
0\\
0\\
1\\
1\\
\end{array} \right ]}^{A_2}
$
\caption{\label{fig:bitsliced}Example of a bit-sliced index~\cite{253268}.}
\end{figure}

The algorithm begins with a Boolean bitwise function that views each of the $N$
input bitmaps as representing a vector of  single-bit numbers.  
Conceptually, these vectors of single-bit numbers
are successively added (pointwise) to an accumulator vector
whose entries  may eventually grow to require
$\Theta(\log N)$ bits each.  The multi-bit accumulator is represented as a 
``bit-sliced index''~\cite{253268}, a collection
of bitmaps $ A_1, A_2, \ldots, A_{\lfloor \log 2N \rfloor}$, where bitmap $A_1$ stores the least-significant bits
of the totals, $A_2$ stores the next-least-significant bits, and so forth (see Fig.~\ref{fig:bitsliced}).  The
totals can be considered to give the bitwise Hamming weight of the inputs; see
Fig.~\ref{fig:hamming-ex}.  We express a Hamming weight using $\lfloor \log 2N \rfloor$~bits, the minimal number of bits required to write $N$ in binary form.

\begin{figure}
\centering
\begin{tabular}{ccc|c|cc}
 \multicolumn{3}{c|}{Inputs} & \multirow{2}{*}{Hamming weight}& \multicolumn{2}{c}{Outputs}\\
  $B_1$  & $B_2$ & $B_3$ &  & $A_2$ & $A_1$\\
 \midrule
  0 & 1& 1& $0+1+1= 2$&1 &0 \\
  0 &0 & 1& $0+0+1=1$&0 &1 \\
  1 &1 & 1& $1+1+1=3$&1 &1 \\
  1 &0 & 0& $1+0+0=1$&0 &1 \\
 \multicolumn{3}{c|}{\vdots} & \vdots&\multicolumn{2}{c}{\vdots}\\
\end{tabular}
\caption{\label{fig:hamming-ex} Computing the bitwise Hamming function.
}
\end{figure}

(Successive addition into an accumulator is not necessarily the best approach to adding 
$N$ 1-bit numbers to obtain $\lfloor \log 2 N \rfloor$-bit Hamming weights.  It is also possible~\cite{symmetric-tr}
to use a balanced binary tree of adders, a ``carry-save'' adder 
approach~\cite{elli:scheduled-vertical-counter}, or (perhaps best) 
a ``sideways-sum'' circuit presented by Knuth~\cite[7.1.2]{KnuthV4A}. 
However, we choose to 
present the approach that most closely resembles the published BSTM algorithm.)

Once we have the Hamming counts, we need to check them to see which
meet threshold $T$.  For this, we can simplify the Range Predicate
computation for bit-sliced indexes, Algorithm~4.2 of O'Neil and
Quass~\cite{253268}.  Rather than check for $\geq T$, we do a
greater-than comparison against $T-1$. 
In Fig.~\ref{fig:hamming-ex}, for $T=2$ we should compute the bitmap 1010\ldots,
since the Hamming counts of the first and third rows exceed $2-1$: $10_2 > 1$,
$01_2 \not > 1$, $11_2 > 1$ and $01_2 \not > 1$. 
 (Again, it is possible to
improve somewhat on the number of bitmap
operations~\cite{symmetric-tr}, but we choose to use the previously
published algorithm, specialized to compute only greater-than.)

The BSTM algorithm is presented in Algorithm~\ref{alg:bstm}.
The correctness of the computations of Hamming counts and greater-than
have been previously established~\cite{rinfret:bit-sliced-arithmetic,253268}.

We can illustrate the algorithm as follows:
\begin{enumerate}
\item Suppose we begin with three bitmaps: $B_1= 0011$,
$B_2= 1010$,
$B_3= 1110$. Before the main loop of the algorithm, we have $A_1= B_1= 0011$ and $A_2=0000$.
\item During the first pass through the main loop ($i=2$), we first compute $C=1010 \land 0011=0010$ and 
$A_1= 1010 \oplus  0011 = 1001$. Because $C$ is not empty, we further need to update $A_2$ to 
$0010$. We now have $j_{\max}=2$.
\item During the second pass through the main loop ($i=3$), we first set $C=1110 \land 1001 = 1000$ and 
$A_1= 1110 \oplus 1001 = 0111$.
Because $C$ is not empty, we have to update $A_2$ to $1000 \oplus  0010 = 1010$.
\item At the end of the main loop, we have $A_1=0111$ and $A_2=1010$ with $j_{\max}=2$.
\item Suppose that the threshold is $T=2$, then the last loop in the algorithm runs from $2$ to $1$. When $j=2$, we set $b_{\mathrm{gt}}=A_2= 1010$ and $b_{\mathrm{eq}}=\neg A_2= 0101$. When $j=1$, we set $b_{\mathrm{eq}}= 0101 \land  0111=0101$. The final answer is $1010$.
\end{enumerate}

\begin{algorithm}
\centering
\begin{algorithmic}[1]
\REQUIRE $N$ bitmaps $B_1, B_2, \ldots, B_N$, a threshold parameter  $T\in \{2,\ldots N-1\}$ 
\STATE // $A$ is the bit-slice-index accumulator for the Hamming weights
\STATE create $\lfloor \log 2N \rfloor$ empty bitmaps $A_1, A_2, \ldots, A_{\lfloor \log 2N \rfloor}$
\STATE $A_1 \leftarrow B_1$
\STATE $j_{\max}\leftarrow 1$
\STATE // keep track of the $A_j$'s being modified
\STATE // Add remaining bitmaps (1-bit numbers) to the accumulator 
\FOR{$i \leftarrow 2$ \textbf{to} $N$}
\STATE  $C \leftarrow B_i \land A_1$;\ \  $A_1 \leftarrow  B_i \oplus A_1$ 
\STATE // Propagate carries (C) to other slices
\STATE $j \leftarrow 2$
\WHILE{$C$ is not empty} 
\STATE  $C , A_j \leftarrow C \land A_j, C \oplus A_j$ 
\STATE $j \leftarrow j+1$
\ENDWHILE
\STATE   $j_{\max}\leftarrow \max (j,j_{\max})$
\ENDFOR
\STATE \label{line:gtr-computation} // Compare Hamming weights against $T-1$
\STATE $b_{\mathrm{eq}} \leftarrow 1111\cdots$
\STATE $b_{\mathrm{gt}} \leftarrow 0000\cdots$
\IF{$j_{\max} < \lfloor \log (2 (T-1)) \rfloor$} 
\RETURN $0000\cdots$
\ENDIF

\FOR{$j \leftarrow j_{\max}$ \textbf{down to} 1}
\IF{bit $j$ is set in $T-1$}
\STATE $b_{\mathrm{eq}} \leftarrow b_{\mathrm{eq}} \land A_j$
\ELSE
\STATE $b_{\mathrm{gt}} \leftarrow b_{\mathrm{gt}} \lor b_{\mathrm{eq}} \land A_j $
\STATE $b_{\mathrm{eq}} \leftarrow b_{\mathrm{eq}} \land \neg A_j$
\ENDIF
\ENDFOR
\STATE return $b_{\mathrm{gt}}$
\end{algorithmic}
\caption{\label{alg:bstm} \bstm\ algorithm.  Each input bitmap $B_i$ is treated as a bit-slice index encoding 1-bit numbers.
}
\end{algorithm}

To analyze the number of bitmap operations, we consider
the following worst-case situation.
The first item occurs in every bitmap and hence has a Hamming count of $N$.
The second item occurs in every bitmap except the first, and in general
the $i^{\mathrm{th}}$ item, for $1 \leq i \leq N$,  occurs in all
bitmaps except for the first $i-1$.
With this worst case,
the first \texttt{for} loop iterates $N-1$ times, doing 2 bitmap operations
before beginning the \texttt{while} loop.  On beginning
 the $i^{\mathrm{th}}$ iteration,
the items have Hamming counts ranging from $0$ to $i-1$; in particular, some
have Hamming counts  with $\Theta(\log i)$
trailing 1s.  Thus
there will be $\Theta(\log i)$ other slices where carry propagation
(involving two operations) is done.  Together, we have 
$2(N-1)+ \sum_{i=1}^{N-1} \Theta(\log i)$ operations to compute
the Hamming weights.
This quantity is $\Theta(N \log N)$, so the number of operations grows
more than linearly in $N$, in the worst case.
There are a few operations required to compare the Hamming weights against $T$.
In the worst case, $j_{\max} = \lfloor \log 2N \rfloor$, and this many iterations 
are done.  Each iteration does 3 bitmap operations (counting ANDNOT as a single
operation), except when a bit of $T-1$ is 1; in that case, only 1 bitmap
operation is done.  If $\#(T-1)$ denotes the Hamming weight of $T-1$, we need
$3 \lfloor \log 2N \rfloor - 2 \#(T-1)$  bitmap operations.  When $N$ is large,
the number of operations for comparison is inconsequential, due to the $\Theta(N \log N)$
worst-case cost to compute Hamming weights.  Nevertheless, this algorithm can
do very few operations in some cases (approximately $2N$
if the maximum Hamming weight is 1 and $T>1$).

An Opt-threshold algorithm can be obtained from a bit-sliced index 
with $O(\log N)$ bitmap operations using ideas from Rinfret et al.~\cite{rinfret:bit-sliced-arithmetic}.

\paragraph{Symmetric functions beyond threshold:} 
We could apply a bit-sliced index to compute
general symmetric functions.
One could use the previous approach to compute the 
$\lfloor \log 2N \rfloor$-bit Hamming weights of
the inputs followed by a 
computation of basic bitmap operations 
for the corresponding test (e.g., is the result between $T_1$ and $T_2$?)
in lieu of the $>$ computation
making up the second half of Algorithm~\ref{alg:bstm}.

In cases where $N$ is small, we are guaranteed to use few operations. Indeed,
Knuth~\cite[7.1.2]{KnuthV4A} observes that since he has
calculated the minimum number of operations (12) to realize any 5-input 
Boolean function, we 
can realize any symmetric Boolean
function of 
$N\leq 31$ inputs using  no more than $12+s(N)$ operations, 
where $s(N) = 5N-2\#(N)-3\lfloor \log N \rfloor -3$ is the number
of operations that a sideways-sum circuit uses to compute the Hamming weight~\cite[Prob. 7.1.2.30]{KnuthV4A}.  
  (For instance,
if $N=31$ we use $5 \times 31 - 2 \times 5 - 3  \times 4 - 3 = 130$ operations 
to compute the Hamming weight; with at most another 12 we can realize \emph{any}
symmetric function.)  

\section{New approaches for Threshold Functions}
\label{sec:newapproaches}
In addition to existing approaches for computing threshold functions,
we also propose a few novel techniques. They can be modified
to solve Opt-threshold queries.

\subsection{Mergeable-count structures.}
\label{sec:mergeable-count-algos}
 
A common approach to computing intersections and unions of several
sets is to do it two sets at a time. To generalize the idea to
symmetric queries, we represent each set as an array of values coupled
with an array of counters. For example, the set $\{1,14,24\}$ becomes
$\{1,14,24\}, \{1,1,1\}$, where the second array indicates the frequency
of each element (respectively). If we are given a second set
($\{14,24,25,32\}$), we supplement it with its own array of counters
$\{1,1,1,1\}$ and can then merge the two: the result is the union of
two sets along with an array of counters ($\{1,14,24,25,32\},
\{1,2,2,1,1\}$). From this final answer, we can deduce both the
intersection and the union, as well as other symmetric operations.

Algorithm \wtwocti\ takes this approach.  Given $N$ input bitmaps, it
orders them by increasing cardinality and then merges each input, starting
with the shortest, into an accumulating total. 
(The merge step is akin to the merge operation in the merge-sort algorithm.)
 A worst-case input
has bitmaps of equal cardinality, each containing $B/N$ items that
are disjoint from any other input.  At the $i^{\textrm{th}}$ step the
accumulating array of counters will have $Bi/N$ entries and this
will dominate the merge cost for the step.  The total time complexity
for this worst-case input is 
$\Theta(\sum_{i=1}^{N-1}Bi/N) = \Theta(BN)$.  For memory use, the same input ends up
growing an accumulating array of counters of size $B$.

Algorithm \wtwocti\ refines this basic 
approach:
although it ends up reading its entire input, during the merging stages 
it can discard elements that cannot achieve the required threshold.
For instance, we can check the accumulating counters
during each merge step.  If there are $i$ inputs left to merge,
then any element that has not achieved a count of at least
$T-i$ can be removed from consideration (``pruned'').  

In large-threshold cases, this pruning is beneficial.
For instance, suppose $T=N-\tau$ for some $\tau \geq 1$.  Any item that has
not occurred in one  of the first $\tau + 1$  bitmaps will be 
pruned.  As these are the smallest bitmaps, they can contain
no more than $(\tau+1)B/N$ items, and this bounds the size of
the accumulator in any of the $N$ merge operations.  The total
cost of the merge operations is thus in $O(B+N (\tau+1)B/N) =
O( \tau B) = O( (N-T) B)$.
 However, pruning 
is mostly unhelpful 
with the worst-case input, if $T=2$.  We cannot discard
any item until the final merge is done, because the last input set
could push the count (currently 1) of any accumulated item to
2, meeting the threshold.  Thus, with $T=2$ we find a worst-case
time bound of $\Omega( (N-T) B)$.

\subsection{\looped\ algorithm}
\label{sec:looped-algo}

Given $N$~bitmaps $B_1, B_2, \ldots, B_N$, the \looped{} algorithm (see Algorithm~\ref{alg:looped})
seeks to compute the threshold problem for all thresholds $1, 2, \ldots, T$ using 
corresponding temporary bitmaps $C_1, C_2, \ldots, C_T$. Let us consider
a concrete example: $B_1=0011$, $B_2=1110$ and $B_3=1000$ with $T=2$. At first, we process
bitmap $B_1$ and get $C_1=0011$, $C_2=0000$. We then
process bitmap $B_2$ and get $C_1=1111$, $C_2=0010$. We
then process the last bitmap to get $C_1=1111$, $C_2=1010$.
As with \bstm, the \looped\ approach also combines basic bitmap operations to
synthesize the threshold operation.

Our algorithm uses dynamic programming and is based on the following recurrence formula: 
$\thresh(T, \{b_1, b_2, \ldots, b_N\}) = \thresh(T, \{b_1, \ldots , b_{N-1}\}) \lor 
   \thresh(T-1,\{b_1, \ldots , b_{N-1}\}) \land b_N$.
I.e., we can achieve a given threshold $T$ over $N$ bits,
either by achieving it over $N-1$ bits, or by having a 1-bit for
$b_N$ and achieving threshold $T-1$ over the remaining $N-1$ bits.
We can  use bit-level parallelism to express this  as a computation over bit vectors;
loops can compute the result specified by the recurrence.  Although 
$\Theta(NT)$ bit-vector operations are used, 
we need only $\Theta(T)$ working bitmaps during the
computation, in addition to our $N$ inputs.

The number of binary bitmap operations is $2NT-N-T^2+T-1$ 
and
depends linearly on $T$, which is unusual compared with
our other algorithms. 
However, the number of bitmap operations is not necessarily a good predictor
 of performance  when using compressed bitmaps. It depends on the dataset. 

\begin{algorithm}
\centering
\begin{algorithmic}[1]
\REQUIRE $N$ bitmaps $B_1, B_2, \ldots, B_N$, a threshold parameter   $T\in \{2,\ldots N-1\}$ 
\STATE create $T$ bitmaps $C_1, C_2, \ldots, C_T$ initialized with false bits
\STATE $C_1 \leftarrow B_1$
\FOR{$i \leftarrow 2$ \textbf{to} $N$}
\FOR{$j \leftarrow \min(T,i)$ \textbf{down to} 2}
\STATE    $C_j \leftarrow C_j \lor (C_{j-1} \land B_i) $
\ENDFOR
\STATE $C_1 \leftarrow C_1 \lor B_i$
\ENDFOR
\STATE return $C_T$
\end{algorithmic}
\caption{\label{alg:looped} \looped\ algorithm.}
\end{algorithm}

An Opt-threshold algorithm is easily obtained from \looped{}: 
first do the calculation with the maximum permitted value of $T$---
i.e., $N$ or $N-1$.  Then find the maximum value $i$ such that
$C_i$ is not empty. 
This algorithm does $\Theta(N^2$) bitmap operations, requiring
$\Theta(N^2 r/W)$ time if we assume bitmap compression is
ineffective.

\subsection{Exploiting run-length coding: \cdom} 
\label{sec:cdom-algo}

Algorithm~\textsc{RunningBitmapMerge} (henceforth \cdom{}) is a refinement of an algorithm presented in  Lemire et al.~\cite{arxiv:0901.3751}.  
The simplest form of the algorithm is for bitmaps that have been
run-length encoded; handling word alignment adds additional complexity
that is discussed in \S~\ref{sec:cdom-algo-ewah}.

See Algorithm~\ref{algo:genrunlengthmultiplefaster} and Fig.~\ref{fig:rbmrg}.  
The approach considers
runs as integer intervals, and
each bitmap provides a sorted sequence
of intervals. For example, the bitmap $B_1=00111000$
might be viewed as the sequence 
(bit: 0, range $[0,1]$; bit 1, range $[2,4]$; bit 0, range $[5,7]$).

Heap $H$ enables us to quickly find, in sorted order, those points where
intervals begin (and the bitmaps involved).  At such points, we calculate the function
on its revised inputs; in the case of symmetric functions such as threshold, this can be quick. 
As we sweep through the data, we update the current count.  Whenever a
new interval of 1s begins, the count increases; whenever a new interval of 0s
begins, the count decreases.   
Assuming $\log N \leq W$,  
the new value of a threshold function can be determined in $\Theta(1)$ time whenever
an interval changes.  (The approach can be used with Boolean functions in general, but 
the complexity analysis might differ.)

Every run passes through a $N$-element heap, giving a running time of
$O(\textsc{RunCount} \log N)$. 
One can implement the $N$ required iterators in
$O(1)$ space each, leaving a memory bound of $O(N)$.

\begin{figure}
\centering
\begin{tabular}{ll|l|l}
              & $B_1$                  & $B_2$                             & Count  \\ \hline
$\rightarrow$ & 0                      & 0                                 & 0\\
$\rightarrow$ & 0                      & \cellcolor[gray]{0.8}1            & 1\\
$\rightarrow$ & \cellcolor[gray]{0.8}1 & \cellcolor[gray]{0.8}1            & 2\\
              & \cellcolor[gray]{0.8}1 & \cellcolor[gray]{0.8}1            & 2\\
              & \cellcolor[gray]{0.8}1 & \cellcolor[gray]{0.8}1            & 2 \\
$\rightarrow$ & 0                      & 0                                 & 0\\
$\rightarrow$ & 0                      & \cellcolor[gray]{0.8}1            & 1\\
$\rightarrow$ & 0                      & \cellcolor[gray]{0.8}1            & 1\\ \hline
\end{tabular}
\caption{\label{fig:rbmrg} Runs, showing positions where new runs begin (and where
the current Hamming-weight count needs to be adjusted).}
\end{figure}

As an extreme example where this approach would
excel, consider a case where each bitmap is either entirely 1s or
entirely 0s.  Then $\textsc{RunCount} = N$, and in $O(N \log N$)
time we can compute the output, regardless of $r$ or $B$.

\subsection{Implementing \cdom{} with EWAH} 
\label{sec:cdom-algo-ewah}

The EWAH implementation of \cdom\  processes runs of clean words as described,
but word alignment means that we must consider dirty words also. 
If the interval from $a'$ to $a$ corresponds to $N_{\textrm{clean}}$ bitmaps with clean runs, of which
$k$ are clean runs of 1s, the implementation distinguishes three  cases:
\begin{enumerate}
\item $T-k \leq 0$: the output is 1, and there is no need to examine the $N  -  N_{\textrm{clean}}$
bitmaps that contain dirty words.  This pruning will help cases when $T$ is small.
\item $T-k >  N  -  N_{\textrm{clean}}$: the output is 0, and there is no need to examine the dirty
words.  This pruning will help cases when $T$ is large.
\item $1 \leq T-k \leq N  -  N_{\textrm{clean}}$: the output will depend on the dirty words.  We can
do a $(T-k)$-threshold over the $N  -  N_{\textrm{clean}}$ bitmaps containing dirty words. 

We process the  $N  -  N_{\textrm{clean}}$ dirty words as follows.
\begin{enumerate}
\item If $T-k=1$ (resp. $T-k = N  -  N_{\textrm{clean}}$), we compute the bitwise OR (resp. AND) between the dirty words. 
\item  If $T-k\geq 128$, we always use 
 \scncnt{} using 64~counters (see \S~\ref{sec:scancount}).
\item  Otherwise, we compute $\beta$, the number of 1s 
in the 
 dirty words. This can be done efficiently in Java 
 since the  \texttt{Long.bitCount} function on desktop processors is typically
 compiled to fast machine code. If $2 \beta \geq (N  -  N_{\textrm{clean}})(T-k)$, 
 we use the \looped{} algorithm (\S~\ref{sec:looped-algo}), otherwise we
 use \scncnt{} again. 
\end{enumerate}
 We arrived at this particular approach by trial and error: 
 we find that it gives reasonable performance.
 
\end{enumerate}

Like \mgopt{} and \dsk{}, \cdom{} has minimal memory usage ($O(N)$, see Table~\ref{tab:complexity-rle-compressed}).
Indeed,  the memory usage of \cdom{} does not depend on the length of the bitmaps ($r$) in contrast to competitive schemes like \scncnt{}, \bstm{} and \looped{}. This might make \cdom{} especially suitable for multicore processing where all cores share the same limited cache memory.

When the bitmaps are poorly compressible, we can view
\cdom{}  as a memory-conscious  version of  \scncnt{}. Indeed, whereas \scncnt{} uses $r$~counters, \cdom{} uses only 64~counters---constantly recycling them.

The algorithm would be a suitable addition to compressed bitmap
index libraries that are RLE-based; as a result of this work,
we have added it to JavaEWAH~\cite{JavaEWAH}---the complete implementation is freely available online.

To illustrate the algorithm, consider the following problem involving 4~bitmaps
and a threshold query with $T=3$.
\begin{enumerate}
\item Without compression, but in
terms of 64-bit words, our 4~bitmaps are \\
$B_1= \{\texttt{\underline{0x0}, 0x0F, \underline{0x00, 0x00, 0x00}, 0x0F, 0x01} \}$,\\
$B_2= \{\texttt{\underline{0x0}, 0xF0F, \underline{0xF$\cdots{}$F, 0xF$\cdots{}$F}, 0x0F, 0x0F, 0x01}\}$ and \\ 
$B_3=B_4=\{\texttt{\underline{0xF$\cdots{}$F, 0xF$\cdots{}$F, 0xF$\cdots{}$F, 0xF$\cdots{}$F}, 0x0F, 0x0F, 0x01}\}$.\\
When using EWAH compression, we have that $B_1$ contains two runs of fill words (containing 0s  and shown underlined) and two runs of dirty words. 
We have that $B_2$ contains two runs of fill words, and two
runs of dirty words, $B_3$ contains one run of fill words and one run of dirty words.
Finally, $B_4$ is identical to $B_3$.
\item The algorithm considers four runs (one for each bitmap). Initially, it considers a run of 0s from $B_1$ (of length 1~word),
a run of 0s from $B_2$ (of length 1~word), and two other runs of 1s (of length 4~words) from $B_3$ and  $B_4$.
Using a heap, it determines that the shortest run has length 1~word. The Hamming weight of the fill words is 2 and there is no dirty word, so immediately it outputs a single fill word of 0s by {case 2}.
\item We  have a run of one dirty word from $B_1$ (\texttt{0x0F}), a run of one dirty word from $B_2$ (\texttt{0xF0F}) 
and the same run of fill words from $B_3$ and $B_4$ (with a remaining length of 3~words). 
Because $T=2$ and we have one fill word made of 1s, the algorithm outputs
the bitwise OR of the two dirty words (\texttt{0xF0F}) by {case 3a}.
\item The algorithm then looks at the beginning of a run of 0s (of length 3~words) in bitmap $B_1$, and at runs of 1s (of length 2~words) in $B_2$, $B_3$ and $B_4$.
The algorithm immediately outputs two fill words of 1s by {case 1}.
\item We have a run of 0s of length 1~word in $B_1$, and  runs of dirty words
from $B_2$,  $B_3$ and $B_4$. The algorithm thus outputs the bitwise AND between the first dirty words from $B_2$,  $B_3$ and $B_4$ (\texttt{0x0F}) by {case 3a}.
\item The algorithm looks at 4~runs of dirty words of length~2 words from $B_1$, $B_2$, $B_3$ and $B_4$. In this instance, case 3c applies. It collects the first 4~dirty words from the 4~bitmaps (\texttt{0x0F}, \texttt{0x0F}, \texttt{0x0F}, \texttt{0x0F}).
The algorithm computes the number of 1s ($\beta = 16$) and it uses the 
 \looped{} algorithm, outputting \texttt{0x0F}. On the next 
 four~dirty words (\texttt{0x01}, \texttt{0x01}, \texttt{0x01}, \texttt{0x01}), it finds that $\beta=4$ and uses
 the \scncnt{} algorithm on the last four dirty words; it outputs \texttt{0x01}.
\item The algorithm concludes with the solution \\
$\{\texttt{0x0, 0xF0F, 0xF$\cdots{}$F, 0xF$\cdots{}$F,0x0F, 0x0F,0x01} \}$. 
\end{enumerate}

\begin{algorithm}
\centering
\begin{algorithmic}
\REQUIRE {$N$ bitmaps $B_1, \ldots, B_N$ over $r$~bits, some Boolean function $\gamma$ such 
as  $\thresh(T, \{\cdot\})$}
\STATE $I_i \leftarrow$ iterator over the runs of identical bits of  $B_i$
\STATE $\Gamma \leftarrow$ a new buffer to store the aggregate of   $B_1, \ldots, B_N$  (initially empty)
\STATE $\gamma \leftarrow$ the bit value determined by $\gamma(I_i,\ldots, I_N)$
\STATE $H \leftarrow$ a new  $N$-element min-heap storing ending values of the runs along with their iterators
\STATE $a' \leftarrow 0$
\WHILE{true}  
\STATE let $a$ be the minimum of all ending values for the runs of $I_1, \ldots, I_N$, determined from $H$
\STATE append run $[a',a]$ to $\Gamma$ with value $\gamma$
\STATE $a' \leftarrow a+1$
\FOR {iterator $I_i$  with a run ending at $a$ (selected from $H$ as root element)}
\STATE increment $I_i$; if $I_i$ has reached the end, terminate the algorithm
\STATE Update $\gamma$ with the new value of $I_i$
\STATE Update the heap $H$ with the new value of $I_i$
\ENDFOR
\ENDWHILE
\end{algorithmic}
\caption{\label{algo:genrunlengthmultiplefaster} Algorithm \cdom.
}
\end{algorithm}

\section{Detailed Experiments}
\label{sec:experiments}

We conducted extensive experiments on the various threshold algorithms,
using EWAH compressed bitmaps generated from real datasets.
The various bitmaps in our study, even within a particular dataset, 
 vary drastically in characteristics such
as density.  
We discuss this in more detail
before giving the experimental results. 

\subsection{Platform} 
Experimental results were gathered on a 
desktop with an Intel Core~i7 2600 (3.4\,GHz, 8\,MB of L3~CPU cache)
 processor with
16\,GB of memory (DDR3-1333 RAM with dual channel). Because all algorithms
are benchmarked after the data has been loaded in memory, disk performance is
irrelevant.

The system ran Ubuntu 12.04LTS with Linux kernel 3.2.
 During experiments,
we disabled dynamic overclocking (Turbo~Boost) and dynamic frequency scaling (SpeedStep).
Software was written in Java (version 1.7), compiled and run using 
OpenJDK (IcedTea 2.4.7) and the OpenJDK 64-bit server JVM\@.


We used the JavaEWAH software library~\cite{JavaEWAH}, version 0.8.1,
for
our EWAH compressed bitmaps. It includes an implementation of the  \cdom{} 
algorithm.
Our measured times were in wall-clock milliseconds.
All our software is single-threaded. 

\subsection{Data}
\label{sec:real-datasets}

Real data tests were done 
with datasets \IMDBthree, 
\PGDVD, \PGDVDtwo, \CensusIncome,  \Tweed\ and \Weather\footnote
{See \url{http://lemire.me/data/symmetric2014.html}.}.
Our first three datasets (\IMDBthree, \PGDVD\ and \PGDVDtwo) are similar
to datasets used in related work~\cite{Li:2008:EMF:1546682.1547171}. They
are \emph{not} indexed as if they were database tables. 
The last three datasets (\Weather, \Tweed\ and \CensusIncome) are more 
representative of content  from relational databases and they are indexed as such (see Fig.~\ref{fig:examplebitmapindex}). 

\IMDBthree\ is
based on descriptions of a dataset used in the work of Li et al.~\cite{Li:2008:EMF:1546682.1547171}, 
in an application looking for actor names that are at a small edit distance from a (possibly
misspelt) name.  
Each bitmap corresponds to a 3-gram found in some actor's name.
The $k^{\mathrm{th}}$ bit in the bitmap indicates whether the  $k^{\mathrm{th}}$ actor's name contains
this 3-gram.

The \PGDVD\ dataset has a bitmap for each of \num{11118} files on the 
Project Gutenberg
DVD~\cite{GutenbergDVD}.  
Each bitmap represents the vocabulary set found in that file
(the total vocabulary had over 2.4~million words).  

\PGDVDtwo\ is similar to \IMDBthree\ 
except that,  
instead of actor
names, we formed $2$-grams from chunks of text from the Project Gutenberg
DVD\@.  Each chunk was obtained by concatenating
paragraphs until we accumulated at least \num{1000} characters.  We rejected
any paragraph with over \num{20000} characters---this protected us from 
some non-text content (e.g.,  the digits of $\pi$) on the Project Gutenberg
DVD\@. 

\begin{table}
\caption{\label{tab:real-data-char}Characteristics of real datasets.
Overall bitmap density is the number of 1s, divided by the product of
the number of rows and the number of bitmaps ($B/(Nr)$).
}
\centering
\tabsize
\setlength{\tabcolsep}{5pt}
{
\begin{tabular} {crrrrrr} \toprule 
Dataset         & \multicolumn{1}{l}{$r$}     & \multicolumn{1}{l}{Attributes} & \multicolumn{1}{l}{Bitmaps} & \multicolumn{3}{c}{\hspace*{5pt}Average Density}\\ 
                &          &            &         &   \multicolumn{1}{r}{\hspace*{10pt}Overall} & \multicolumn{1}{r}{In M-C workload} & \multicolumn{1}{r}{In Sim workload} \\
\midrule
\IMDBthree      & 1783816  &    ---        &  50663  &  $4.1\times 10^{-4}$ &   --- \hspace{2em}                  & $1.9\times 10^{-2}$  \\
\PGDVD          & 2439448  &    ---        & 11118   &  $2.9\times 10^{-4}$ &  --- \hspace{2em}                 & $3.7\times 10^{-3}$  \\
\PGDVDtwo       & 3513575  &     ---       &  755    &  $2.8\times 10^{-1}$ &  --- \hspace{2em}                 & $6.1\times 10^{-1}$  \\  \\
\CensusIncome   & 199523   & 42         & 103419  &  $4.1\times 10^{-4}$ & $1.5\times 10^{-1}$ & $3.4\times 10^{-1}$  \\ 
\Tweed          & 11245    & 53         & 1167    &  $4.5\times 10^{-2}$ & $2.0\times 10^{-1}$ & $5.5\times 10^{-1}$  \\
\Weather        & 1015367  & 19         & 18647   &  $1.0\times 10^{-3}$ & $7.6\times 10^{-2}$ & $1.2\times 10^{-1}$  \\ \bottomrule
\end{tabular}
}
%
%
%


\end{table}

We also chose three more conventional datasets in the context of relational databases.  Two have  many attributes, 
\CensusIncome~\cite{arxiv:0901.3751,MLRepository} and \Tweed~\cite{tweed}. 
The former is a census extract; the latter is a small dataset containing
historical information on terrorist attacks in Europe,
for which we used all attributes, rather than the projection used by
Webb et al.~\cite{webb2013}. 
We also used the entries for September 1985
of a larger dataset (\Weather)~\cite{hahn:weatherbench,304214}.  This particular month has been used previously~\cite{304214},
although
the previous use had projected only 9~attributes, whereas we used all of them.
We selected just one month of data because the full dataset (123~million rows) 
caused several of the tested algorithms 
to run out of memory.  
It would have been difficult to report meaningful aggregate 
results with such failures. 

A bitmap index was built for each conventional dataset, and it had a 
bitmap for every attribute value.  Row reordering can
improve
 RLE-compressed bitmap indexes~\cite{arxiv:0901.3751}, but
it is not always possible.  Our indexes used the given (unsorted) row order.
In \CensusIncome, one attribute is
responsible for \num{99800} of the bitmaps; the remaining \num{3619}~bitmaps are
much denser than these \num{99800}.  Together, our real datasets cover a
range of application areas, lengths, widths and densities.

By design, our work does not consider external-memory indexes on very large datasets. Thus our datasets are chosen
so the bitmaps for each query 
fit in RAM\@. However, even if our machines had more than 16\,GB of RAM, we might still want to partition the problems so that bitmaps do not span much more than a few million bits, to alleviate caching issues.

\subsection{Queries Used}
\label{sec:queries}

To assess our algorithms, we generated two random workloads, one with 
\num{5000}~Many-Criteria queries and the other
with  \num{5000}~Similarity queries (see \S~\ref{sec:advancedqueries}). 
\begin{itemize}

\item To generate a Many-Criteria query, we randomly chose a dataset.
Many-Criteria queries do not make much
sense for \IMDBthree, \PGDVD\ or \PGDVDtwo.  For instance, almost all
3-grams have extremely sparse bitmaps and empty results can be
expected, even with $N$ large and $T$ small.  Therefore, we chose the
dataset with equal probability from \{\CensusIncome,~\Tweed,~\Weather\}.  
Having determined the dataset, we chose $N$ next.  
We could pick random values of $N$ uniformly at random in a range ($[3,1000]$), but most values of $N$ would then be relative large ($\gg 10$). Instead, 
we used
a discretized log-uniform distribution with 
$\log N\sim U[\log 3,\log 1000]$, which
resulted in a workload where small values of $N$
were more common, but large values of $N$ sometimes occurred. 
We then chose (uniformly at random, with replacement) $N$
attributes on which criteria were established, by choosing one of
their bitmaps uniformly.  We finally randomly chose an integer threshold
$T$ uniformly from $[2,N'-1]$, where $N'$ is the number of attributes
on which criteria had been established.
\item We considered Similarity
queries with $n$ prototypes (henceforth Similarity($n$)).
For such a  query,  we selected (with equal
probability) one of our datasets.  Then we chose $n$~distinct prototypes $\{r_i
\,|\, i \leq n\}$, each represented by a row identifier chosen uniformly from
$[0,r)$.
 We then
found the set of bitmaps matching at least one of them,
$\bigcup\{ B_i \,|\, \exists j \mbox{\ such that\ } B_i[r_j]=1\}$; 
we have that $N$ was the number
of matching bitmaps.

The probability of
Similarity(1), Similarity(5), Similarity(10), Similarity(15) and
Similarity(20) queries were \SI{20}{\percent} each.
\end{itemize}

In the last columns of Table~\ref{tab:real-data-char}, we present the average density of the bitmaps involved in our query workloads. 

We agree with Jia et al.~\cite{jia2012eti} that it does not make sense
to time queries whose answers are empty.  Regardless whether we had a
Many-Criteria or a Similarity query, if the answer to the threshold
query was empty and $T>2$, we chose (uniformly at random) a new value
of $T$ between 2 and the existing value of $T$.  If the threshold
query had an empty answer when $T=2$, we discarded the query and generated
a new one.

Considering the \num{10000} queries in the two workloads, there were \num{54} 
queries with $N>1000$:
the maximum value of $N$ was \num{11115} 
whereas the
average $N$ was \num{165}.  The maximum value of $T$ was \num{10863}, 
 but the average was \num{42}.  
The largest set of input data, in terms of storage, was \SI{185}{MB};
in terms of cardinality, it was 740\, million items.


The bitmaps involved in our queries are denser than the average bitmap. Indeed, the last three columns
in Table~\ref{tab:real-data-char} differ: the first shows the average density
of bitmaps from the dataset, whereas the second and third respectively show the average density of the
bitmaps actually selected in our two workloads.
We see that the latter are denser (anywhere from twice as dense to 1000~times denser). This is a consequence of 
how we pick the queries.
\begin{itemize}
\item  
Many-Criteria queries
tend to choose dense bitmaps because the sparsest bitmaps frequently come from the same (high cardinality)
attribute, and all attributes are given an equal probability. 
\item   
For Similarity queries, 
we note that   
denser bitmaps are more likely to appear in 
$\bigcup\{ B_i \,|\, \exists j \mbox{\ such that\ } B_i[r_j]=1\}$.
\end{itemize}

\paragraph{Choosing $\mu$ for \dsk:} \label{sec:pickmu}
The \dsk\ algorithm requires a tuning parameter $\mu$, which depends on the dataset.  
Li et al.~\cite{Li:2008:EMF:1546682.1547171}
sketch a process for choosing $\mu$: 
\begin{itemize}
\item For each dataset, select a representative workload of queries.
\item For each query, execute \dsk\ with various choices of $\mu$, 
recording the $\mu$ that produced the
fastest answer for that query.
\item Average the recorded $\mu$ values for a dataset.
\end{itemize}

We followed their approach.
For the workload, we generated
\num{500} queries using the random query generation process already described
for Many-Criteria queries.
We tried up to \num{20} values of $\mu$ for each query, using the relationship
$L=T/(\mu \log M +1)$ given by Li et al.\ ($M$ is the cardinality of the largest bitmap)
to choose $\mu$ values.
When $T \leq 20$, we tried $L=1, L=2, \ldots, L=T-1$.
  Otherwise, we tried all values in 
$ [T-5,T) \cup \left \{ \left \lceil \frac {T-6}{15} i \right \rceil \,|\,i \in [1,15] \right \}$.  
For 
\CensusIncome, \Tweed\ and \Weather, the respective $\mu$ values were 
\num{0.0388},  \num{0.0452} and \num{0.0444}.
We then repeated the process with \num{500} Similarity queries, obtaining
$\mu$ values of
\num{0.180} (\IMDBthree), \num{0.0752} (\PGDVD), \num{0.00481} (\PGDVDtwo)
\num{0.0560} (\CensusIncome), \num{0.0112} (\Tweed) and \num{0.0351} (\Weather).

\paragraph{Competitions:} We assess the effectiveness of the various
algorithms by measuring their wall-clock times on the queries in our workload.
Each query 
can be viewed as a competition between algorithms.

Unfortunately, for some of the larger queries, \wtwocti\ (\S~\ref{sec:mergeable-count-algos}) was not able to
complete without running out of memory.  
It is unfair to give the algorithm a nearly infinite running time when trying to compute
its aggregate performance over the workload.  However, it is also unfair
to omit the running time from an average, as it is excusing a result where even a 
good algorithm would take a long time.  Our solution is to assign the running time
of the slowest algorithm that \emph{did} complete the competition.  

\subsection{Experimental Effects of $N$ and $T$}

As previewed in Table~\ref{tab:complexity-rle-compressed},
our theoretical bounds suggest that the various algorithms' running times are all 
affected\footnote{For table entries 
(such as that for \scncnt ) where $B$ is given but $N$ is not explicit, note that
$B$ grows as $N$ grows: given a set of $N$ bitmaps with $B$ 1s, if a new non-empty bitmap
is added, the total number of 1s increases.} by $N$.
Some algorithms are affected by $T$ and others are highly sensitive to the
characteristics of the datasets being processed.
A few anecdotal examples given here illustrate these effects and
help confirm/augment our theoretical bounds, as well give 
some idea of the constants that are abstracted away during our asymptotic analyses. (See \cite{symmetric-tr} for more extensive experiments.) 

\begin{figure}
\begin{centering}
\subfloat[\cdom, \scncnt, \bstm\ and \looped]{\includegraphics[width=0.45\textwidth]{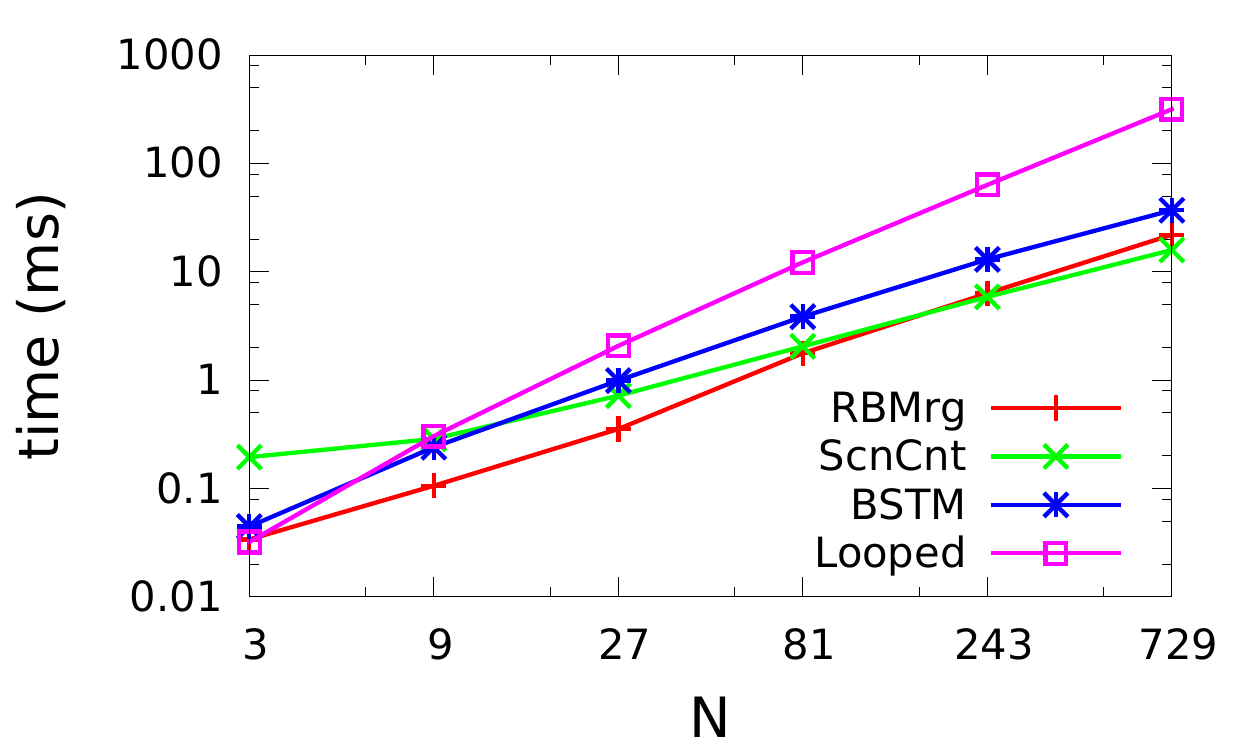}
}
\subfloat[\cdom, \dsk, \wtwocti\ and \mgopt]{\includegraphics[width=0.45\textwidth]{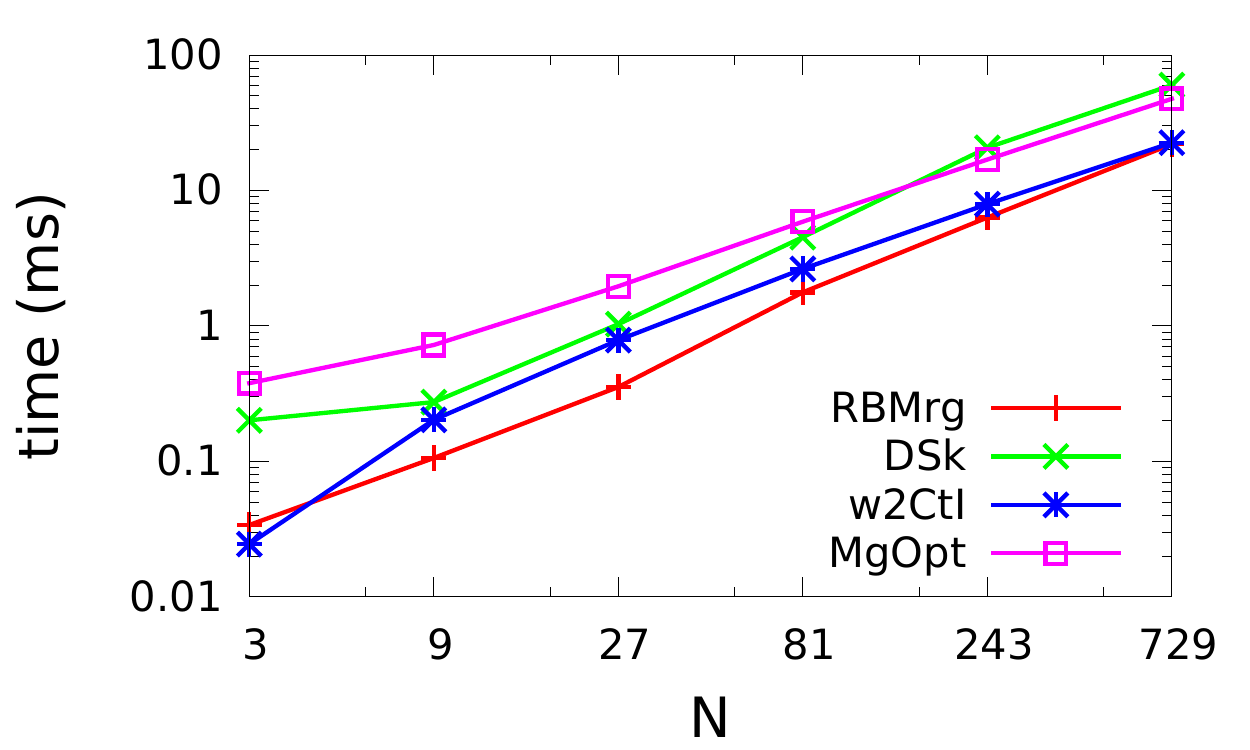}
}
\caption{\label{fig:timesVsN}Effect of $N$ on the running times of the algorithms, for Many-Criteria majority queries on \CensusIncome.
For clarity, we use two plots to represent the 7~algorithms: the same \cdom{} timings appear in the two plots.
}
\end{centering}
\end{figure}

We first fixed the dataset (\CensusIncome) and kind of query (Many-Criteria majority) to examine the effect of $N$
in that particular scenario.  
For chosen values of $N$, we took 100~queries on 
\CensusIncome\  and, for each algorithm, averaged
their running times.  These are majority queries:
threshold queries  with $T = \lceil N/2 \rceil$ and $N$ odd.
(Unlike our normal workload, we had a mixture of queries returning
empty and non-empty results.)
In this scenario, we have that $r$ and $W$ are fixed while $B$, $B'$, $T$ and $N-T$ grow with $N$.
From Table~\ref{tab:complexity-rle-compressed}, we might expect (using an admittedly na\"ive analysis) the  
running time of \scncnt{} to grow linearly ($N$), 
the running time of
\mgopt{}, \bstm{},
\dsk{} and  
\cdom{} to grow as $N \log N$ and the running time of 
\wtwocti{} and \looped{} to grow quadratically ($N^2$).
(Experimentally, it is often difficult to distinguish linear growth from $N \log N$, but 
quadratic growth will stand out as having a larger slope on a log-log plot.)
Figure~\ref{fig:timesVsN}
shows how our algorithms behaved in this particular test as $N$ was
changed.  

Focusing on $N \geq 9$, we see that \cdom\ had the best absolute performance except when $N$ was very large. In such cases, \scncnt\ was faster.
In terms of
growth rate (corresponding to slope in a log-log plot), 
\looped\ stands out, with a growth rate that corresponds to approximately
$N^{\log_3 5} \approx N^{1.5}$ --- better than our $O(N^2)$ worst-case bound
suggests, but still worse than the other algorithms.  The slopes of \cdom\
and \dsk\  are higher than those of the other algorithms.  In other tests
we rarely see
\wtwocti\ performing well, due to its large memory requirement.  However, on
these queries against our small \CensusIncome\ dataset,
it seemed to display a slightly sub-linear running time growth in $N$, far
better than our quadratic bound indicated.
\scncnt\ was similar in apparently having sub-linear growth.
The query-generation approach means that $B$, the number of set bits, grows
proportionally with $N$.  
However, the total number of possible items, $r$, is constant and, for
this dataset, we have $r$ approximately 8~times larger than the number
of set bits in an average input bitmap.  With $N=3$, we have $B$ significantly
less than $r$.  They are comparable at $N=9$.  This can explain apparently
sublinear growth as $N$ grew from 3 to moderately large values of $N$.
The explanation for \wtwocti\ appears simpler: on this dataset,
our majority queries had empty answers for $N>27$.  The ever-larger
thresholds presented more opportunities for pruning that \wtwocti\ 
exploited.  We might have expected similar improvements from both
\mgopt\ and \dsk, but only \mgopt\ seems to have had them.  The reason
may be that \dsk\ prunes especially well when $N-T$ is small. Since
we have $T\approx \frac N 2$, we see \dsk\ outperforming \mgopt\ for
small $N$, but then becoming closer to \mgopt\ as $N$ increases.

Discussion during our theoretical analyses has indicated that large $T$ values improve
pruning possibilities---and hence should lead to improved running times---for
several algorithms (\wtwocti, \mgopt, \dsk\ and \cdom).  However, these pruning effects are 
data dependent and
hence, except for \wtwocti, were not reflected in  our asymptotic bounds.
(In fact, our bounds for \mgopt\ and \dsk\  actually suggest running time might
\emph{increase} somewhat with $T$.)
Our \looped\ algorithm is expected to grow linearly with $T$, due to a $NT$ term.
Moreover, small values of $T$ can lead to pruning in $\cdom$.
Experiments can help us see, at least in specific cases, the effects of
pruning that are not captured by our asymptotic running-time bounds.

\begin{figure}
\begin{centering}
\includegraphics[width=0.95\textwidth]{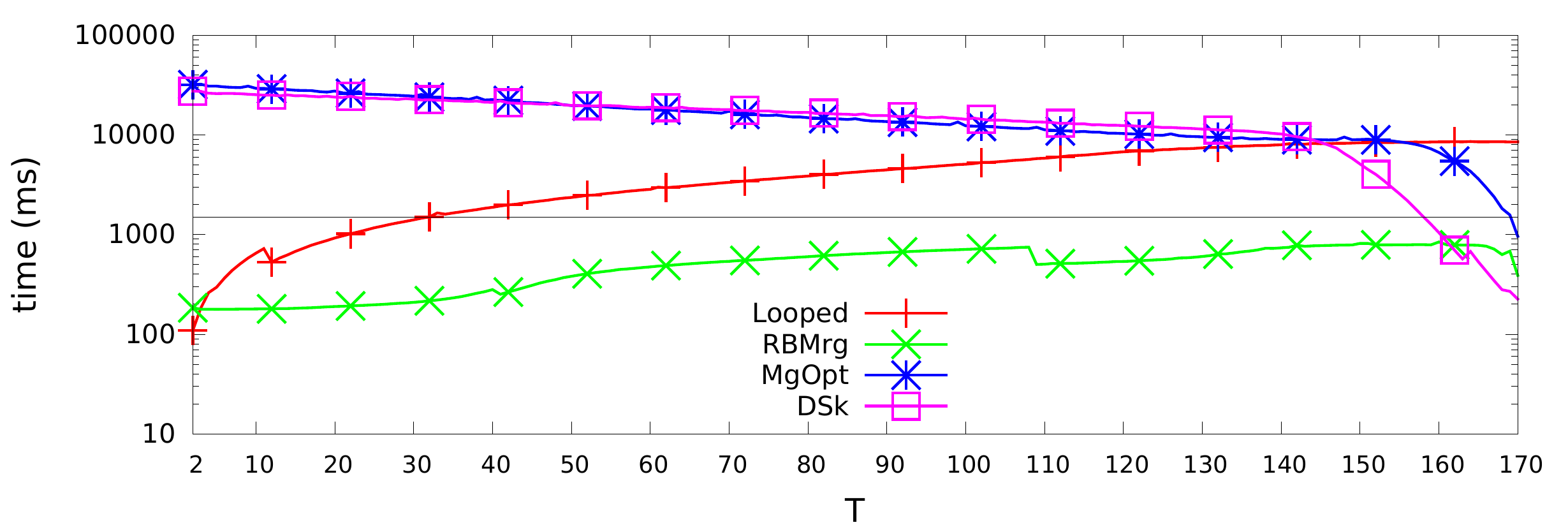}
\caption{\label{fig:timesVsT}Effect of $T$ on the running times of several algorithms.
 $N=171$, and the dataset
is \PGDVDtwo. Other algorithms were less affected by $T$. The \scncnt\ 
and \bstm\ algorithms took about \SI{1500}{ms} for all values of
$T$.
The \wtwocti\ algorithm
dropped steadily from about \SI{13000}{ms} for small $T$ to about 
\SI{10000}{ms} for the largest $T$ values. 
}
\end{centering}
\end{figure}

We then chose an arbitrary query (a Similarity(1) query against
\PGDVDtwo): 
Fig.~\ref{fig:timesVsT} shows the effect of varying $T$, on one particular 
set of \num{171} bitmaps.
Absolute times are shown, but on a logarithmic 
scale. 
Increasing $T$ (and thereby decreasing the size of the answer)
affected algorithms differently, and
\dsk\ is particularly notable,
improving two orders of magnitude and going from one of the worst
algorithms for small $T$, to the best for large $T$.  
It is difficult to see, but \cdom\ had a \SI{43}{\percent} 
speedup  when $T$
increased from 169 to 170. Overall, it tended to perform best when $T$
was small, however.
The different pruning opportunities can affect which algorithm is
fastest for a given $T$.  
For this collection of
bitmaps, we got best results from \looped\ at $T=2$, then
\cdom\  until $T \approx 160$, after which \dsk\ was fastest.
The potentially enhanced pruning of \dsk\ over \mgopt\ was not manifest until
$T=145$ on this dataset, whereas for Fig.~\ref{fig:timesVsN} even
majority queries usually showed an advantage for \dsk.  

\subsection{Comparing Algorithms on Our Workloads}

Because the state of our system varies slightly over time, 
we make an error when measuring the time required by  the implementation of an algorithm.  We think of  the \emph{true} performance
of the implementation of the algorithm as its best possible speed on
a given query.
We can measure this best possible speed with little error by repeating the execution hundreds of times. However, given our \num{10000}~queries and 7~algorithms, these repeated tests would require more than a year to complete.
Thus we tested each algorithm on each query only a few times.

Moreover, when comparing algorithms, we did not merely want to decide whether an
algorithm is superior to another; for this purpose a standard
statistical test would have sufficed. Instead we wanted to compare the
results numerically, so we first estimated our measurement error 
by generating 5~random queries. 
Because timings errors are always additive, given a query, we ran each
algorithm on each query 200~times. The minimum timing is assumed to
be error-free:  the fastest test out of 200~tests is 
a good approximation of the fastest possible result. 
Any larger timing is in error. We have 6~datasets
and 7~algorithms, so we collected $200 \times 6 \times 7$~measurement
errors per query. We found that the 99$^{\mathrm{th}}$-percentile
error was less than \SI{10}{\percent} for the 5~reference queries.
 
We then considered our set of \num{10000}~queries.
For each query, dataset and algorithm, we fixed the number of repetitions so
that the total running time is at least
\SI{1}{\second}.  
Supposing the measured running times on
some query are 
$t_1$ and $t_2$
for two given algorithms, we say that the first algorithm is faster
only if $t_1 < 0.8 \times t_2$.  We anticipate mis-identifying a
superior algorithm less than \SI{1}{\percent} of the time.

Tables~\ref{tbl:beats-matrix-sim}~and~\ref{tbl:beats-matrix-multi} compare
 each pair of algorithms using our two workloads. The
cell associated with the row for algorithm $\mathcal{A}_1$ and the
column for algorithm $\mathcal{A}_2$ gives the number of times that
$\mathcal{A}_1$ had performance superior to that of $\mathcal{A}_2$.

For all cases when $t_1 \leq t_2$, we record the percentage
improvement measured. (A percentage improvement of $x$ means that
$\mathcal{A}_1$ is $1/(1-x)$~times faster than $\mathcal{A}_2$. That
is, improvements of \SI{99}{\percent}, \SI{90}{\percent},
\SI{80}{\percent}, \SI{50}{\percent} indicate that we have
$100\times$, $10\times$, $5\times$ and $2\times$ the speed.)
To assess these performance improvements (ignoring the
possibility of measurement error), we show the 
time reductions
that could be obtained for the query, by using $\mathcal{A}_1$ instead of
$\mathcal{A}_2$.  We show the 
$50^{\mathrm{th}}$- and $75^{\mathrm{th}}$-percentile and maximum time reductions in percentage. 

We round percentage reductions down, thus
percentage reductions of 99~mean speedups of
\emph{at least} 2~orders of magnitude are possible by switching algorithms.
Note that even the weakest algorithm outperforms each of the others
(excepting \cdom),
even if rarely.  The final column in the table shows the number of workload
queries where the row's algorithm was the best (ignoring
possible measurement error).  We see that results
are similar on the two workloads, and  the superior algorithms are
\cdom\ (\SI{80}{\percent} of the queries), \scncnt\ 
(\SI{15}{\percent}).

The final row represents the case where an oracle picks the fastest algorithm for each query. 
As expected, because \cdom{} is best 
about \SI{80}{\percent} of the time, the median of the percentage improvements is zero for this algorithm. 
The final row shows that \emph{every} algorithm performs badly
on at least one instance (e.g., \cdom{} is beaten by \SI{80}{\percent} once, which means that another algorithm is $5\times$ faster). We see that \scncnt, \dsk, \looped, \mgopt\ and \wtwocti\ are sometimes
at least two orders of magnitude slower than necessary. 
At the $75^{\textrm{th}}$
percentile level  \cdom\ is the clear winner (which is expected, given that
it is best \SI{80}{\percent} of the time). 
  \scncnt\ and \bstm\ 
are similar : each is typically about five times slower than the
best algorithm.
Although \scncnt\ is the fastest algorithm 
at least \SI{15}{\percent} of the time versus \SI{0}{\percent}
for \bstm, the comparison may not seem so lopsided when we consider
that \bstm\ was clearly superior to \scncnt\ more than \SI{20}{\percent}
of the time.  

\begin{table}
\caption{\label{tbl:beats-matrix-sim}
Percentage of competitions (Similarity workload) 
where the row's algorithm was at least \SI{20}{\percent} faster than the column's algorithm, and
beneath it, the percentage improvements from the row's algorithm.
 We show
the median, $75^{\textrm{th}}$-percentile, and maximum percentage improvement. 
An improvement of \SI{99}{\percent} means at least $100\times$ speed. \\[1ex] 
The final column shows the percentage of cases where the row's algorithm was measured to be fastest. 
}
\centering\setlength{\tabcolsep}{4pt}
\tabsize
\begin{tabular}{cp{.090\textwidth}p{.090\textwidth}p{.090\textwidth}p{.090\textwidth}p{.090\textwidth}p{.090\textwidth}p{.090\textwidth}p{.090\textwidth}} \toprule
vs  & \textsc{RBMrg}  & \textsc{ScnCnt}  & \textsc{Looped}  & \textsc{DSk}  & \textsc{w2CtI}  & \textsc{BSTM}  & \textsc{MgOpt}  & fastest \\ \midrule
\textsc{RBMrg} & & \parbox{.085\textwidth}{76\,\%{\footnotesize \\73 86 99}}& \parbox{.085\textwidth}{96\,\%{\footnotesize \\86 91 99}}& \parbox{.085\textwidth}{94\,\%{\footnotesize \\91 97 99}}& \parbox{.085\textwidth}{99\,\%{\footnotesize \\93 97 99}}& \parbox{.085\textwidth}{100\,\%{\footnotesize \\75 81 98}}& \parbox{.085\textwidth}{98\,\%{\footnotesize \\91 96 99}}& \parbox{.085\textwidth}{80\,\%}\\[2.5ex]
\textsc{ScnCnt} & \parbox{.085\textwidth}{12\,\%{\footnotesize \\56 69 80}}& & \parbox{.085\textwidth}{73\,\%{\footnotesize \\72 88 99}}& \parbox{.085\textwidth}{77\,\%{\footnotesize \\86 90 96}}& \parbox{.085\textwidth}{90\,\%{\footnotesize \\83 87 97}}& \parbox{.085\textwidth}{58\,\%{\footnotesize \\52 68 96}}& \parbox{.085\textwidth}{82\,\%{\footnotesize \\85 90 98}}& \parbox{.085\textwidth}{15\,\%}\\[2.5ex]
\textsc{Looped} & \parbox{.085\textwidth}{2\,\%{\footnotesize \\30 46 66}}& \parbox{.085\textwidth}{19\,\%{\footnotesize \\62 82 99}}& & \parbox{.085\textwidth}{54\,\%{\footnotesize \\74 90 99}}& \parbox{.085\textwidth}{62\,\%{\footnotesize \\65 83 99}}& \parbox{.085\textwidth}{17\,\%{\footnotesize \\38 68 96}}& \parbox{.085\textwidth}{51\,\%{\footnotesize \\69 89 99}}& \parbox{.085\textwidth}{3\,\%}\\[2.5ex]
\textsc{DSk} & \parbox{.085\textwidth}{1\,\%{\footnotesize \\17 39 64}}& \parbox{.085\textwidth}{17\,\%{\footnotesize \\57 74 92}}& \parbox{.085\textwidth}{38\,\%{\footnotesize \\72 86 99}}& & \parbox{.085\textwidth}{31\,\%{\footnotesize \\32 66 99}}& \parbox{.085\textwidth}{24\,\%{\footnotesize \\50 71 94}}& \parbox{.085\textwidth}{22\,\%{\footnotesize \\21 40 96}}& \parbox{.085\textwidth}{3\,\%}\\[2.5ex]
\textsc{w2CtI} & \parbox{.085\textwidth}{0\,\%{\footnotesize \\13 17 30}}& \parbox{.085\textwidth}{8\,\%{\footnotesize \\60 75 92}}& \parbox{.085\textwidth}{27\,\%{\footnotesize \\54 72 99}}& \parbox{.085\textwidth}{36\,\%{\footnotesize \\25 36 69}}& & \parbox{.085\textwidth}{14\,\%{\footnotesize \\35 51 89}}& \parbox{.085\textwidth}{29\,\%{\footnotesize \\22 36 90}}& \parbox{.085\textwidth}{0\,\%}\\[2.5ex]
\textsc{BSTM} & \parbox{.085\textwidth}{0\,\%}& \parbox{.085\textwidth}{21\,\%{\footnotesize \\28 38 99}}& \parbox{.085\textwidth}{70\,\%{\footnotesize \\53 67 99}}& \parbox{.085\textwidth}{64\,\%{\footnotesize \\83 89 96}}& \parbox{.085\textwidth}{76\,\%{\footnotesize \\80 87 98}}& & \parbox{.085\textwidth}{71\,\%{\footnotesize \\78 87 95}}& \parbox{.085\textwidth}{0\,\%}\\[2.5ex]
\textsc{MgOpt} & \parbox{.085\textwidth}{0\,\%{\footnotesize \\7 12 29}}& \parbox{.085\textwidth}{13\,\%{\footnotesize \\58 74 93}}& \parbox{.085\textwidth}{37\,\%{\footnotesize \\58 76 99}}& \parbox{.085\textwidth}{21\,\%{\footnotesize \\15 23 49}}& \parbox{.085\textwidth}{27\,\%{\footnotesize \\25 52 98}}& \parbox{.085\textwidth}{15\,\%{\footnotesize \\38 64 88}}& & \parbox{.085\textwidth}{0\,\%}\\[2.5ex]
fastest & \parbox{.085\textwidth}{{\ }{\footnotesize \\0 0 80}}& \parbox{.085\textwidth}{{\ }{\footnotesize \\65 84 99}}& \parbox{.085\textwidth}{{\ }{\footnotesize \\87 92 99}}& \parbox{.085\textwidth}{{\ }{\footnotesize \\90 97 99}}& \parbox{.085\textwidth}{{\ }{\footnotesize \\93 97 99}}& \parbox{.085\textwidth}{{\ }{\footnotesize \\77 83 98}}& \parbox{.085\textwidth}{{\ }{\footnotesize \\91 96 99}}& \\[2.5ex]

\bottomrule
\end{tabular}
\end{table}

\begin{table}
\caption{\label{tbl:beats-matrix-multi}
Results on the Many-Criteria workload, in the same format as Table~\ref{tbl:beats-matrix-sim}.}
\centering\setlength{\tabcolsep}{4pt}
\tabsize
\begin{tabular}{cp{.090\textwidth}p{.090\textwidth}p{.090\textwidth}p{.090\textwidth}p{.090\textwidth}p{.090\textwidth}p{.090\textwidth}p{.090\textwidth}} \toprule
vs  & \textsc{RBMrg}  & \textsc{ScnCnt}  & \textsc{Looped}  & \textsc{DSk}  & \textsc{w2CtI}  & \textsc{BSTM}  & \textsc{MgOpt}  & fastest \\ \midrule
\textsc{RBMrg} & & \parbox{.085\textwidth}{75\,\%{\footnotesize \\66 80 99}}& \parbox{.085\textwidth}{91\,\%{\footnotesize \\75 86 98}}& \parbox{.085\textwidth}{99\,\%{\footnotesize \\88 94 99}}& \parbox{.085\textwidth}{98\,\%{\footnotesize \\89 95 99}}& \parbox{.085\textwidth}{99\,\%{\footnotesize \\66 79 99}}& \parbox{.085\textwidth}{99\,\%{\footnotesize \\86 93 99}}& \parbox{.085\textwidth}{77\,\%}\\[2.5ex]
\textsc{ScnCnt} & \parbox{.085\textwidth}{9\,\%{\footnotesize \\20 29 48}}& & \parbox{.085\textwidth}{56\,\%{\footnotesize \\72 84 96}}& \parbox{.085\textwidth}{82\,\%{\footnotesize \\85 89 96}}& \parbox{.085\textwidth}{95\,\%{\footnotesize \\80 83 87}}& \parbox{.085\textwidth}{63\,\%{\footnotesize \\51 59 75}}& \parbox{.085\textwidth}{82\,\%{\footnotesize \\82 86 95}}& \parbox{.085\textwidth}{18\,\%}\\[2.5ex]
\textsc{Looped} & \parbox{.085\textwidth}{3\,\%{\footnotesize \\26 51 76}}& \parbox{.085\textwidth}{34\,\%{\footnotesize \\58 76 98}}& & \parbox{.085\textwidth}{65\,\%{\footnotesize \\70 89 99}}& \parbox{.085\textwidth}{70\,\%{\footnotesize \\71 87 99}}& \parbox{.085\textwidth}{33\,\%{\footnotesize \\33 58 93}}& \parbox{.085\textwidth}{61\,\%{\footnotesize \\64 86 99}}& \parbox{.085\textwidth}{5\,\%}\\[2.5ex]
\textsc{DSk} & \parbox{.085\textwidth}{0\,\%{\footnotesize \\5 13 16}}& \parbox{.085\textwidth}{12\,\%{\footnotesize \\57 77 98}}& \parbox{.085\textwidth}{19\,\%{\footnotesize \\35 53 89}}& & \parbox{.085\textwidth}{24\,\%{\footnotesize \\49 76 99}}& \parbox{.085\textwidth}{9\,\%{\footnotesize \\37 54 87}}& \parbox{.085\textwidth}{7\,\%{\footnotesize \\15 26 65}}& \parbox{.085\textwidth}{0\,\%}\\[2.5ex]
\textsc{w2CtI} & \parbox{.085\textwidth}{0\,\%{\footnotesize \\16 30 57}}& \parbox{.085\textwidth}{4\,\%{\footnotesize \\47 68 97}}& \parbox{.085\textwidth}{16\,\%{\footnotesize \\33 50 83}}& \parbox{.085\textwidth}{50\,\%{\footnotesize \\30 42 77}}& & \parbox{.085\textwidth}{2\,\%{\footnotesize \\36 55 86}}& \parbox{.085\textwidth}{23\,\%{\footnotesize \\16 30 70}}& \parbox{.085\textwidth}{1\,\%}\\[2.5ex]
\textsc{BSTM} & \parbox{.085\textwidth}{0\,\%{\footnotesize \\2 6 11}}& \parbox{.085\textwidth}{23\,\%{\footnotesize \\42 62 98}}& \parbox{.085\textwidth}{45\,\%{\footnotesize \\49 66 91}}& \parbox{.085\textwidth}{83\,\%{\footnotesize \\72 83 99}}& \parbox{.085\textwidth}{95\,\%{\footnotesize \\67 76 98}}& & \parbox{.085\textwidth}{84\,\%{\footnotesize \\65 76 98}}& \parbox{.085\textwidth}{0\,\%}\\[2.5ex]
\textsc{MgOpt} & \parbox{.085\textwidth}{0\,\%{\footnotesize \\7 14 28}}& \parbox{.085\textwidth}{13\,\%{\footnotesize \\54 78 98}}& \parbox{.085\textwidth}{21\,\%{\footnotesize \\33 51 84}}& \parbox{.085\textwidth}{41\,\%{\footnotesize \\19 27 63}}& \parbox{.085\textwidth}{30\,\%{\footnotesize \\32 65 99}}& \parbox{.085\textwidth}{8\,\%{\footnotesize \\36 55 85}}& & \parbox{.085\textwidth}{0\,\%}\\[2.5ex]
fastest & \parbox{.085\textwidth}{{\ }{\footnotesize \\0 0 76}}& \parbox{.085\textwidth}{{\ }{\footnotesize \\59 78 99}}& \parbox{.085\textwidth}{{\ }{\footnotesize \\74 86 98}}& \parbox{.085\textwidth}{{\ }{\footnotesize \\89 95 99}}& \parbox{.085\textwidth}{{\ }{\footnotesize \\89 95 99}}& \parbox{.085\textwidth}{{\ }{\footnotesize \\67 80 99}}& \parbox{.085\textwidth}{{\ }{\footnotesize \\87 93 99}}& \\[2.5ex]

\bottomrule
\end{tabular}
\end{table}

\paragraph{Beating \scncnt :}  In \S~\ref{sec:with-and-without-index} we  
suggested that \scncnt\ could be beaten; indeed, we can see this
by inspecting the \scncnt\ column in 
Tables~\ref{tbl:beats-matrix-sim}~and~\ref{tbl:beats-matrix-multi}.
To be more precise, our workloads contained a query that, 
compared to \scncnt,  was 
answered $1100\times$ faster using \cdom, 
another query that was also $1100\times$ faster with \looped,
one that was $300\times$ faster with \bstm,   
one that was $70\times$ faster using \dsk, 
one that was $34\times$ faster with  \wtwocti, 
and 
one where \mgopt\ was $81\times$ faster.
These extreme cases involve 3~datasets with long bitmaps ($r$ is large)
 and queries
involving a few especially sparse input bitmaps 
($N \leq 4$ and $B$ is small)---conditions especially
difficult for \scncnt.  At least in such cases,
\scncnt\ can be beaten by orders of magnitude.



\subsection{Performance Across Workload Subsets}
\label{sec:workload-performance}

Table~\ref{tab:mixed-workload-table} shows the total time taken by each 
algorithm across both workloads, or across a portion of the workload(s)
meeting certain criteria shown in the first column.
(Since {Tables~\ref{tbl:beats-matrix-sim}~and~\ref{tbl:beats-matrix-multi} 
showed such similar results, we combine the two workload into an overall
composite workload.)
Table~\ref{tab:mixed-workload-table} shows the effect of large $N$,
small $T$ or $N$, the kind of query, or the dataset.

\begin{table}
\caption{\label{tab:mixed-workload-table} Total time to process queries of various groups. The top line of each group is the total time. Then four lines 
give the $25^{\textrm{th}}$-, $50^{\textrm{th}}$-, $75^{\textrm{th}}$-percentile, and maximum query times. Values for \cdom\ are absolute (seconds for total
time; ms for percentile values).  Values for all other algorithms are relative---
the measured time has been normalized by dividing it by the corresponding time for \cdom.
}
\centering\setlength{\tabcolsep}{4pt}
\tabsize
\begin{tabular}{cp{.095\textwidth}p{.095\textwidth}p{.07\textwidth}p{.07\textwidth}p{.07\textwidth}p{.07\textwidth}p{.07\textwidth}} \toprule
\input{aggregate-table.tex}
\bottomrule
\end{tabular}
\end{table}

The table shows
that \looped, \dsk, \mgopt\ and \wtwocti\ can have some
extremely expensive queries, although fewer than \SI{25}{\percent} of the queries are
extremely expensive.   The apparent preference for \cdom\ toward
the top of the table partly breaks down when we examine individual datasets at
the bottom of the table.   
The large size of \PGDVDtwo\ and
the excellent performance of \cdom\ 
 on this large dataset act together to
dominate the overall results. 
Also, results are dominated by
larger values of $N$, despite our generating workloads so that small-$N$
queries were more frequent than large-$N$ ones.
For instance, in all the cases where \scncnt\ did best overall (Many-Criteria
queries, \IMDBthree, \PGDVD\ and \Weather), note that
\cdom\ significantly outperformed \scncnt\ at the median level.
As well, costs were
dominated by Similarity queries; while equal in number to Many-Criteria
queries, they included the queries with the largest values of $N$.

To visualize or aggregate this data, we should consider that the
workload involves datasets of widely different size: there are three
orders of magnitude difference between the total volume of data for our \Tweed\
queries and our \PGDVDtwo\ queries.
Instead of merely timing the queries, we measure their
throughput:  
amount of input data divided by the time necessary to complete the query,
expressing the result in MB/s. 
Given an algorithm and a dataset, we use the harmonic mean to 
obtain an aggregate throughput value.  However, for display purposes
it is convenient to show the reciprocal throughput.  For instance,
the stacked bar charts in Fig.~\ref{fig:mixed-workload-barchart} can be
viewed as representing  times (in seconds) on some hypothetical
workload in which 1\,MB 
of bitmap data had been processed by the queries
for each dataset.  For our workload, \cdom, \scncnt\ and
\bstm\ are strongly preferred to the others.   
Figure~\ref{fig:mixed-workload-barchart-bykind} shows that,
for our relational datasets---the only ones
that were used with both Many-Criteria and Similarity 
queries---\cdom\ is the clear winner.

\newcommand{\h}{${\mathrm H}$}
\newcommand{\hopt}{${\mathrm H}_{\mathrm{opt}}$}
\newcommand{\hds}{${\mathrm H}_{\mathrm{ds}}$}

\begin{figure}
\subfloat [Full y range]{\includegraphics[width=.5\textwidth]{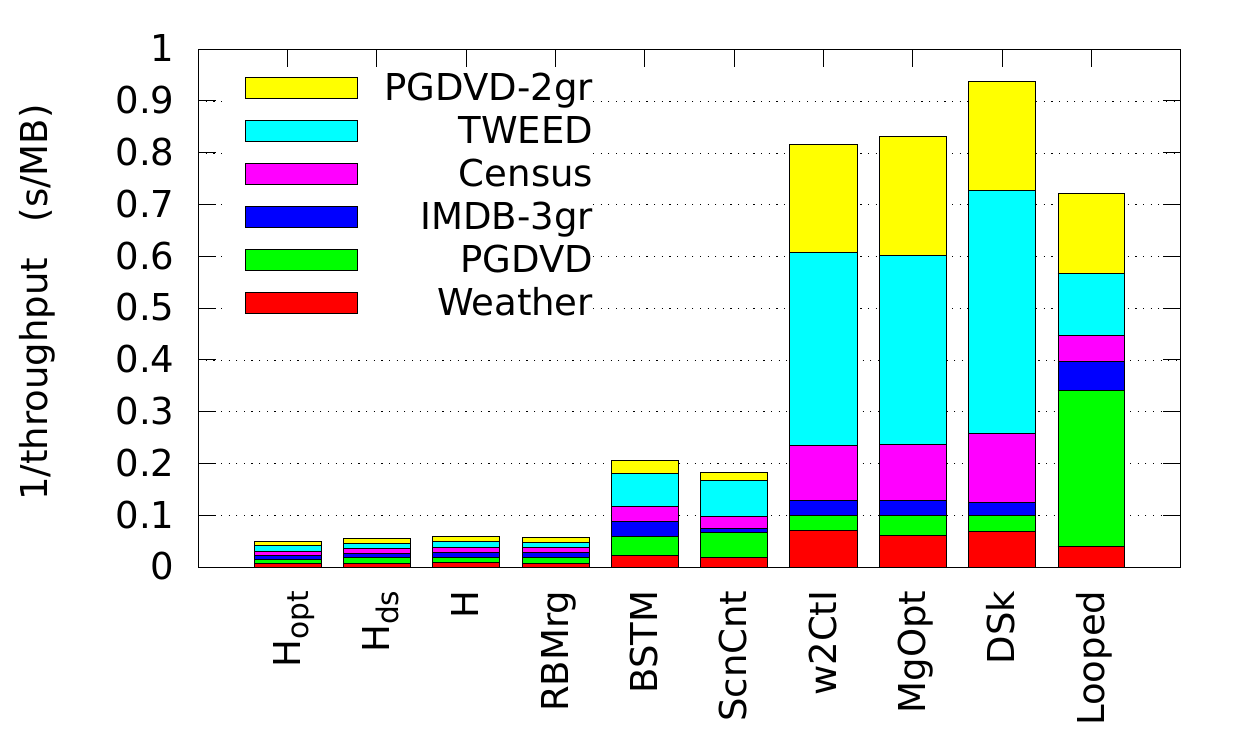}}
\subfloat [Reduced y]{\includegraphics[width=.5\textwidth]{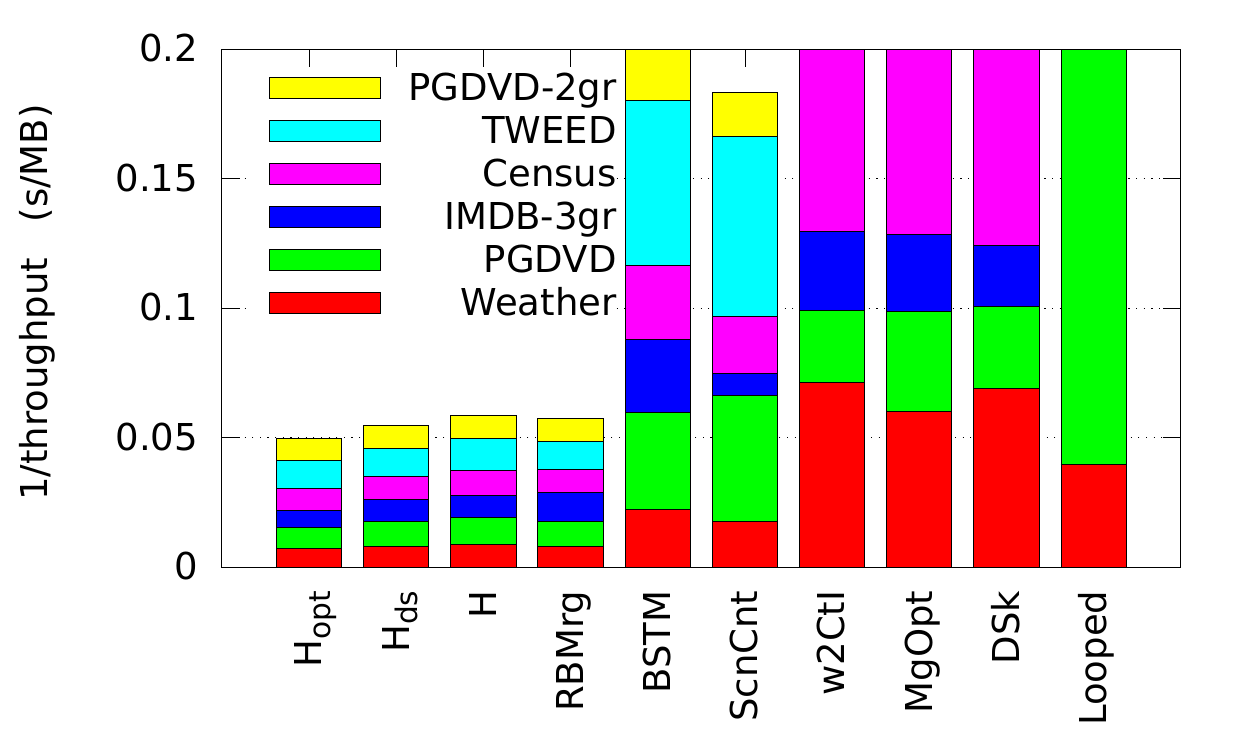}}
\caption{\label{fig:mixed-workload-barchart} 
Aggregate throughput on each dataset. Bar height represents
 the number of seconds
for
a workload containing \SI{1}{MB} of bitmap 
data from each dataset.
\hopt, \hds\ and \h\ are discussed in
\S~\ref{sec:hybrid}.
}
\end{figure}
 
\begin{figure}
\subfloat [Similarity]{\includegraphics[width=.5\textwidth]{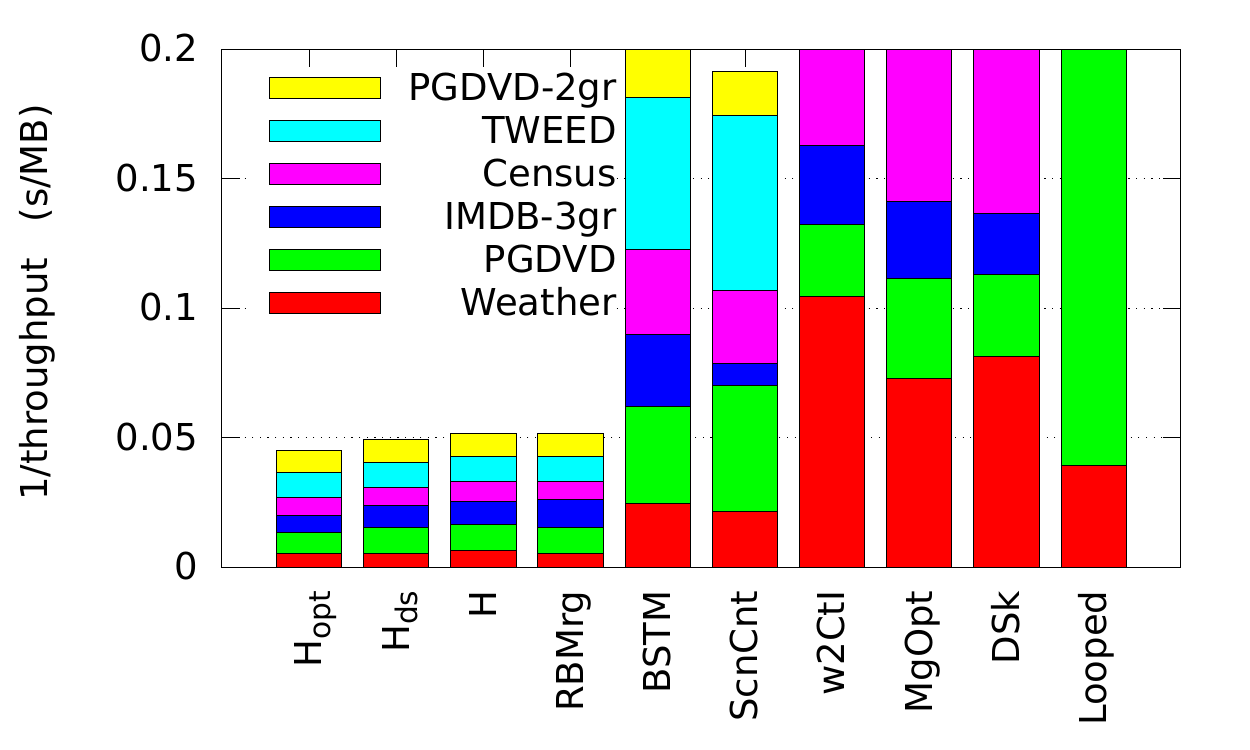}}
\subfloat [\label{subfig:many-crit-barchart}Many Criteria]{\includegraphics[width=.5\textwidth]{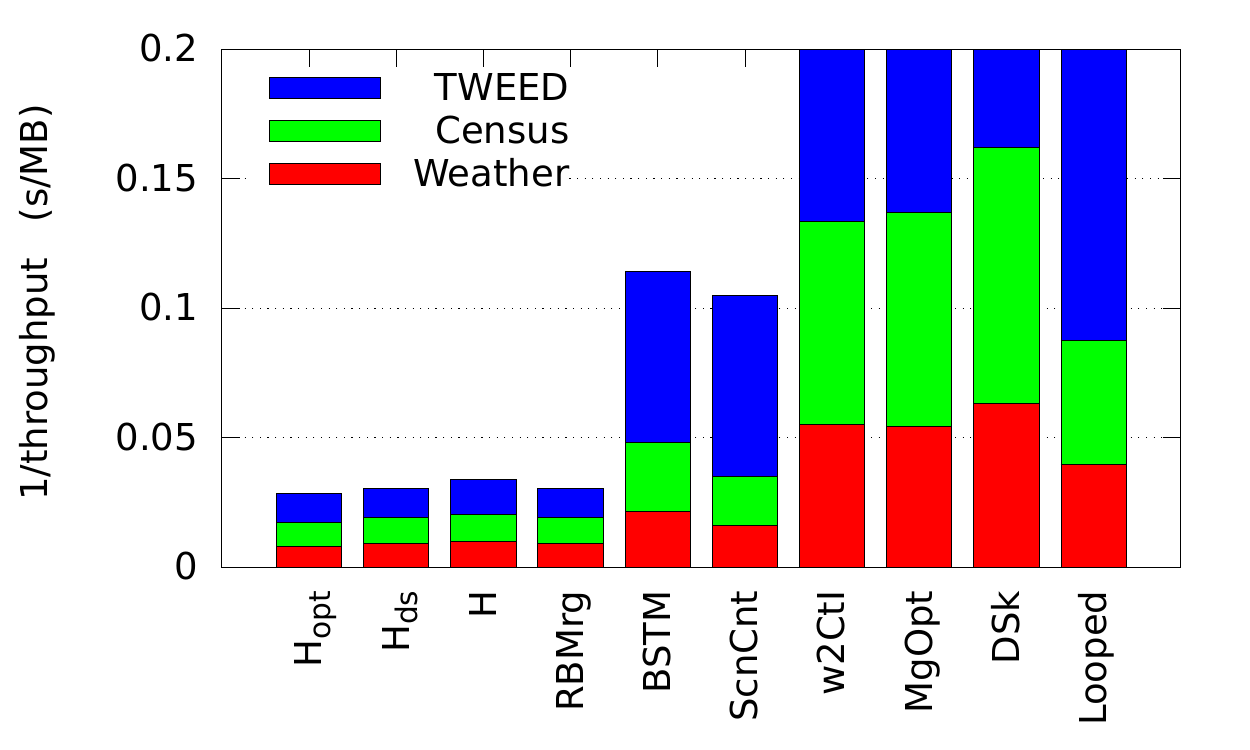}}
\caption{\label{fig:mixed-workload-barchart-bykind} \cdom\ excels
in both workloads, but for the Similarity workload, \scncnt\ does
a better job on \IMDBthree.
}
\end{figure}

In several applications we can expect that $N$ will not be
particularly large, or that typical queries will usually have $T
\approx N$.  Figure~\ref{fig:special-queries-in-workload} shows the
results for these cases.  
On such queries, while \cdom\ is best, \bstm\ 
is better than \scncnt: 
for a typical query with $N$ small, the cost of initializing and
scanning the $r$ counters is being amortized over a small volume of
bitmap data.
We also see that \looped\ is a viable algorithm for $N \leq 16$.

Figure~\ref{fig:special-queries-in-workload} also shows the situation
for the workload queries where $T \geq 0.75 N$.  This situation is one
where pruning-based algorithms such as \wtwocti, \mgopt\ and \dsk\
can excel.  Indeed, we see them doing  well on \IMDBthree,
\Weather\ and (for \wtwocti\ and \dsk) \PGDVD\@.  However, there are other
datasets where they still perform badly.  
Altogether, on workloads similar to
ours, the benefits from pruning do not seem to be worth the risks of
using these algorithms.  Nevertheless, there are some applications
where the requirement for a large $T$ can be met. For instance, using
the formula of
Sarawagi and Kirpal~\cite{Sarawagi:2004:ESJ:1007568.1007652}
with strings of length $N=64$, if we are interested in finding the strings of
edit distance at most two from some target using trigrams, the
appropriate threshold is $T=64+3-1-2\times 3=60$.

\begin{figure}
\subfloat[\label{subfig:n-under-seventeen}Queries with $N \leq 16$]{\includegraphics[width=.5\textwidth]{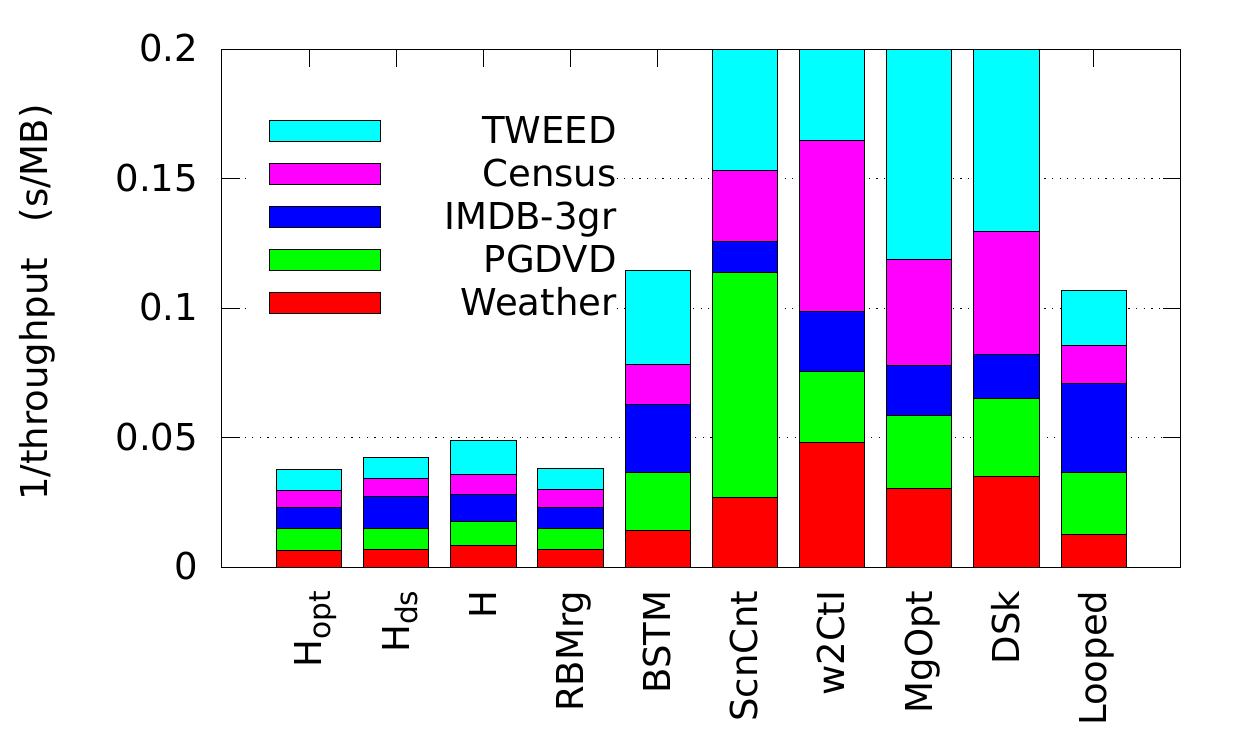}}
\subfloat[Queries with $T \geq .75N$]{\includegraphics[width=.5\textwidth]{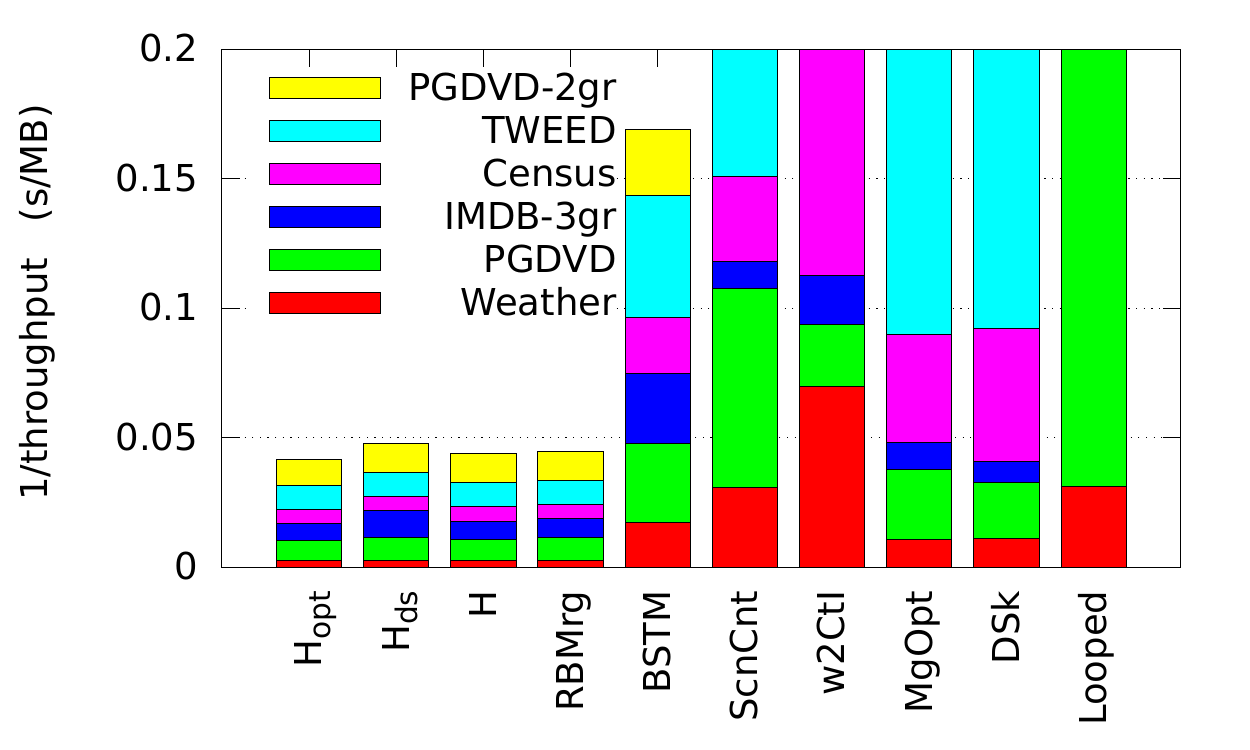}}
\caption{\label{fig:special-queries-in-workload} Normalized workload times for
queries with small $N$ and for queries with $T \geq .75 N$. }
\end{figure}

\subsubsection{The advantage of \cdom.}
Our results show that \cdom\ was usually the fastest algorithm,
especially over datasets coming from relational tables and for
Many-Criteria queries.
This speed advantage is due to fewer executed instructions, rather than
cache effects: experiments showed
that the processor executed about two instructions
per cycle (IPC) for  all implementations. (We saw 1.9~IPC for \scncnt;
2.0 for \wtwocti;
2.1 for \bstm, 
\looped\ and \cdom; and
2.2 for \mgopt\ and \dsk.)

One reason for the advantage of \cdom\ is that it ends up solving 
a threshold problem over the
dirty words, and our implementation adaptively  switches
between algorithms \looped\ and \scncnt.  In essence, it gets a benefit
from RLE encoding, and then combines the strengths of two other
efficient algorithms.  An initial implementation had done a na\"ive
computation (iterating over all bit positions) and this implementation
of \cdom\ was usually not  competitive with \scncnt\ or  \bstm.
Solving the threshold subproblem effectively on the dirty words was
crucial, and 
 our hybrid of \scncnt\ and \looped\ made
the revised implementation fast.

\section{Hybrid Algorithms}
Our success in handling dirty words
adaptively suggests that an adaptive, hybrid approach might also
be a better way to solve threshold problems on compressed bitmaps. 

\subsection{An Execution-Time Model}
\label{sec:model}

\begin{table}
\caption{\label{tbl:running-time-estimates} Running time estimates for good algorithms (this excludes \wtwocti, \mgopt\ and \dsk ).}
\centering
\begin{tabular}{llp{6cm}} \toprule
Algorithm         &  Time complexity estimate  & \hfill Fitted coefficients \\ \midrule
\scncnt           &  $c_{\mathrm{sc},1}\times r+c_{\mathrm{sc},2}\times B$ & \hfill $c_{\mathrm{sc},1}=  2.072\times 10^{-5}\pm 7.6\times10^{-7}$\\ 
                  &                                                          &\hfill $c_{\mathrm{sc},2}=2.683\times10^{-6}\pm 6.1\times10^{-9}$  \\   
\looped           & $c_{\mathrm{\looped}}\times T \times \textsc{EWAHSize}$    &\hfill  $c_{\mathrm{\looped}}= 1.306\times 10^{-6}\pm 2.9\times 10^{-9}$ \\    
\bstm             & $c_{\mathrm{\bstm}}\times \textsc{EWAHSize} \times \ln N$  & \hfill $c_{\mathrm{\bstm}}=3.133\times 10^{-5} \pm 1.6 \times 10^{-7}$\\
\cdom             & $c_{\mathrm{\cdom}}\times \textsc{EWAHSize} \times \ln N$  & \hfill $c_{\mathrm{\cdom}}=1.592\times 10^{-6} \pm 5.3 \times 10^{-9}$ \\    
\bottomrule
\end{tabular}
\end{table}

To guide an adaptive algorithm,  we need to estimate the running times
of the more promising algorithms, in terms of the
limited data that a DBMS might
be expected to maintain.

Table~\ref{tbl:running-time-estimates} shows estimates of the running-time functions
over our workload.  
They were derived by least-squares fitting 
our running-time bounds  in \S~\ref{sec:existingapproaches}--\ref{sec:newapproaches}  
and Table~\ref{tab:complexity-rle-compressed} to 
the measured times for the competitions in the workload.
To account for bitmap compression, we substitute \textsc{EWAHSize} where
 $Nr/W$ occurred in Table~\ref{tab:complexity-rle-compressed}. 
Given a bound of  $O(f(x_1, x_2, \ldots, x_k)$), we modeled
the running time as
$c f(x_1, x_2, \ldots, x_k)$ and fitted $c$ according to the measured
running times.  Algorithm \scncnt\ had two independent terms and we used a 
separate constant for each term.
Also, for \cdom, we felt it would be unreasonable to expect 
the \textsc{RunCount}
of the bitmaps to be cataloged.  Instead, we used \textsc{EWAHSize} as a proxy.

 Our time-complexity estimates for \bstm, \looped\ and \cdom\ are shown
against the actual data in 
Fig.~\ref{fig:bstm-fit-quality}.
We see the fits are not particularly good, but we seldom underestimate running
times by more than a factor of 2.  (Our overestimates are frequently off
by larger factors.)  This may be good enough to avoid selecting 
an algorithm that is badly suited for a query.

\begin{figure}
\centering
\subfloat [\bstm]{\includegraphics[width=0.3\textwidth]{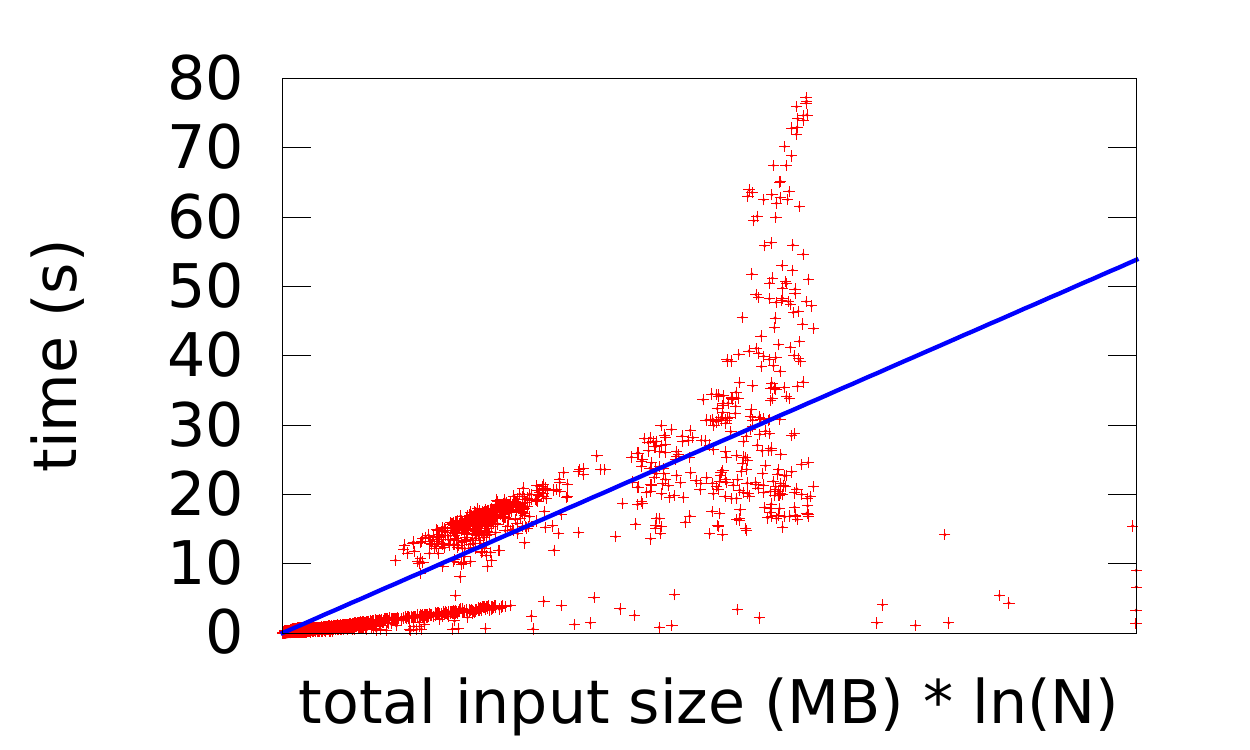}}
\subfloat [\looped]{\includegraphics[width=0.3\textwidth]{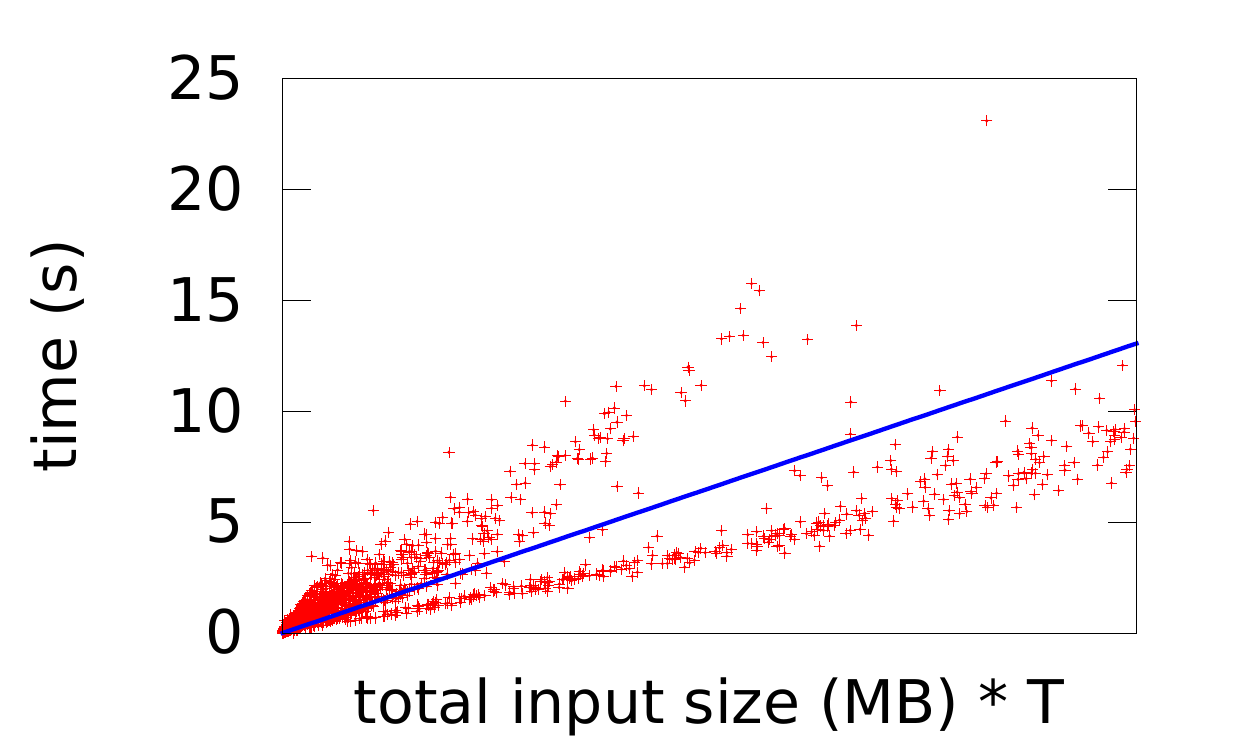}}
\subfloat [\cdom]{\includegraphics[width=0.3\textwidth]{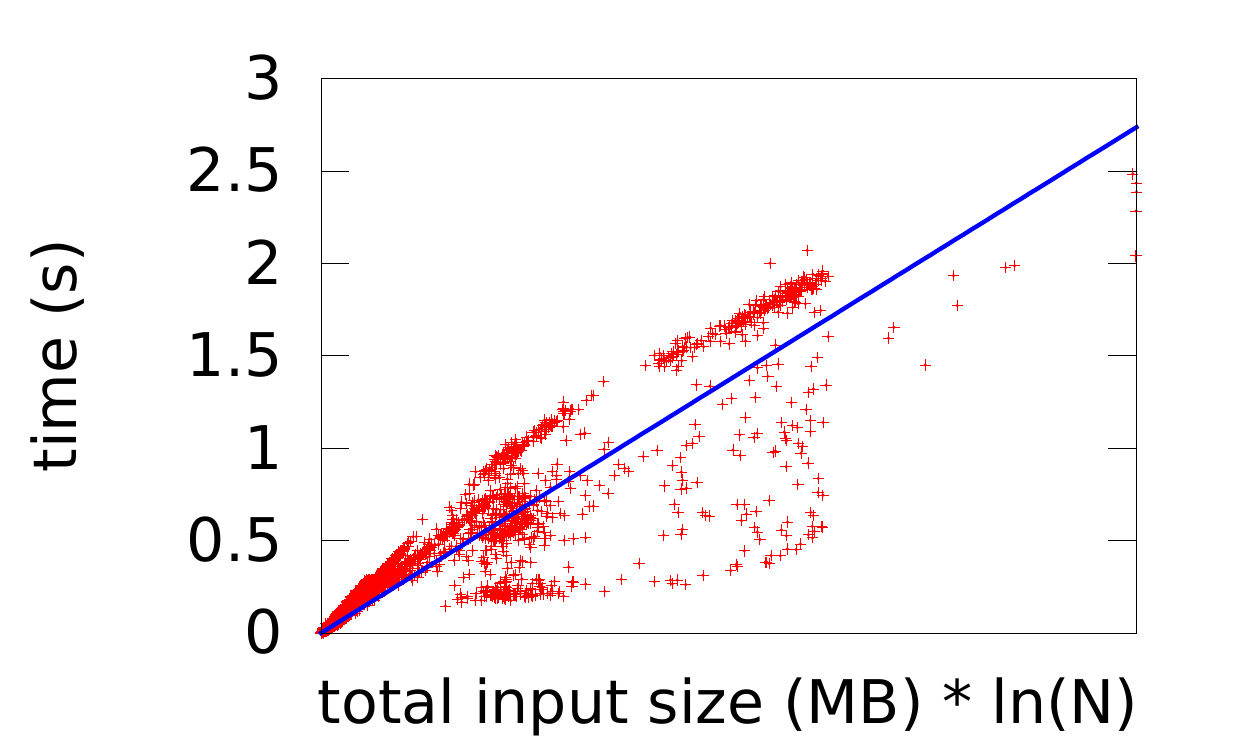}}
\caption{\label{fig:bstm-fit-quality}The running times for \bstm\ and \cdom\ depend on the total compressed 
size of the input bitmaps and a $\log N$ factor. 
The running time for \looped\ depends on the total compressed size and $T$.
We show least-squares
lines (passing through the origin) to fit these models.
}
\end{figure}

\subsection{Algorithms}
\label{sec:hybrid}
We experimented with hybrid algorithms \h\ and \hds, described below.
For comparison purposes, \hopt\ is the hybrid algorithm that always
chooses the fastest algorithm for any query.

\paragraph{Hybrid algorithm, \h.}

Since we have multiple alternative algorithms for the same problem,
there are sophisticated approaches for choosing the best algorithm 
for a given
application~\cite{Hoos:2012:PO:2076450.2076469}.  However, we can get reasonably good
performance with two simple approaches for choosing the appropriate
algorithm.

Our first approach, hybrid algorithm \h{}, evaluates the running-time estimates
given in Table~\ref{tbl:running-time-estimates}.  It then selects the algorithm
predicted to run fastest.  Mathematically, the estimate for
\bstm\ is about 20 times larger than the estimate for \cdom, so
we should never choose \bstm.  
Algebraic manipulation of the time estimates for \cdom\ and \looped\
shows that when  $T < \frac{c_{\cdom}}{c_{\looped}} \ln N$ we should choose
\looped\ in preference to \cdom.  Conveniently, one does not need to know \textsc{EWAHSize}.
 
A weakness of this approach is that it is based explicitly
on the performance of our particular test computers.
While slightly inaccurate estimates
may not lead to bad decisions, those using this approach on systems that differ
significantly should conduct their own benchmarks 
and adjust the coefficients.

\paragraph{Adjust-by-dataset algorithm, \hds.}
Faced with a collection of queries over disparate datasets, an obvious approach
is to select the algorithm entirely on the basis of the dataset.  Perhaps some
initial profile runs would be used to select the algorithm to be used consistently
on a dataset.  This \hds\ approach was tested; on our
workloads we used \scncnt\ for all queries against \IMDBthree,
and \cdom\ for all other queries.  (We chose this combination
by inspecting Fig.~\ref{fig:mixed-workload-barchart}.)

\paragraph{Optimal hybrid algorithm, ${\mathrm H}_{\mathrm{opt}}$.}

For comparison purposes, we can determine the effect of the optimal hybrid 
algorithm, \hopt,
which always selects the best algorithm for any competition as an oracle would.  Since we have already run
every algorithm during the competition, this is easy for us to do.  Of course,
in practice one would
not have this information---\hopt\ exists only to make comparisons.

\subsection{Evaluation of Hybrid Algorithms}

As  Figs.~\ref{fig:mixed-workload-barchart}--\ref{fig:special-queries-in-workload}
show, sometimes there is little reason to rely on anything other than
\cdom, and choosing otherwise only hurts performance.  
Yet Fig.~\ref{fig:mixed-workload-barchart-hhelps} shows two other cases
where hybrid approaches helped. On queries with large $N$, 
\h\ had a \SI{28}{\percent} improvement
over \cdom. (\hopt\ and \hds\ had respective improvements over \cdom\ 
of \SI{31}{\percent} and \SI{13}{\percent}.)  For our text-derived datasets
the improvements
were smaller: \SI{4}{\percent}, \SI{9}{\percent} and \SI{23}{\percent}
respectively for \h, \hds\ and \hopt.  Looking at Figs.~\ref{subfig:many-crit-barchart}~and~\ref{subfig:n-under-seventeen} we see cases where \h\ and \hds\ incorrectly
chose to use algorithms other than \cdom, leading to slightly worse results.


\begin{figure}
\subfloat [$N \geq 200$]{\includegraphics[width=.5\textwidth]{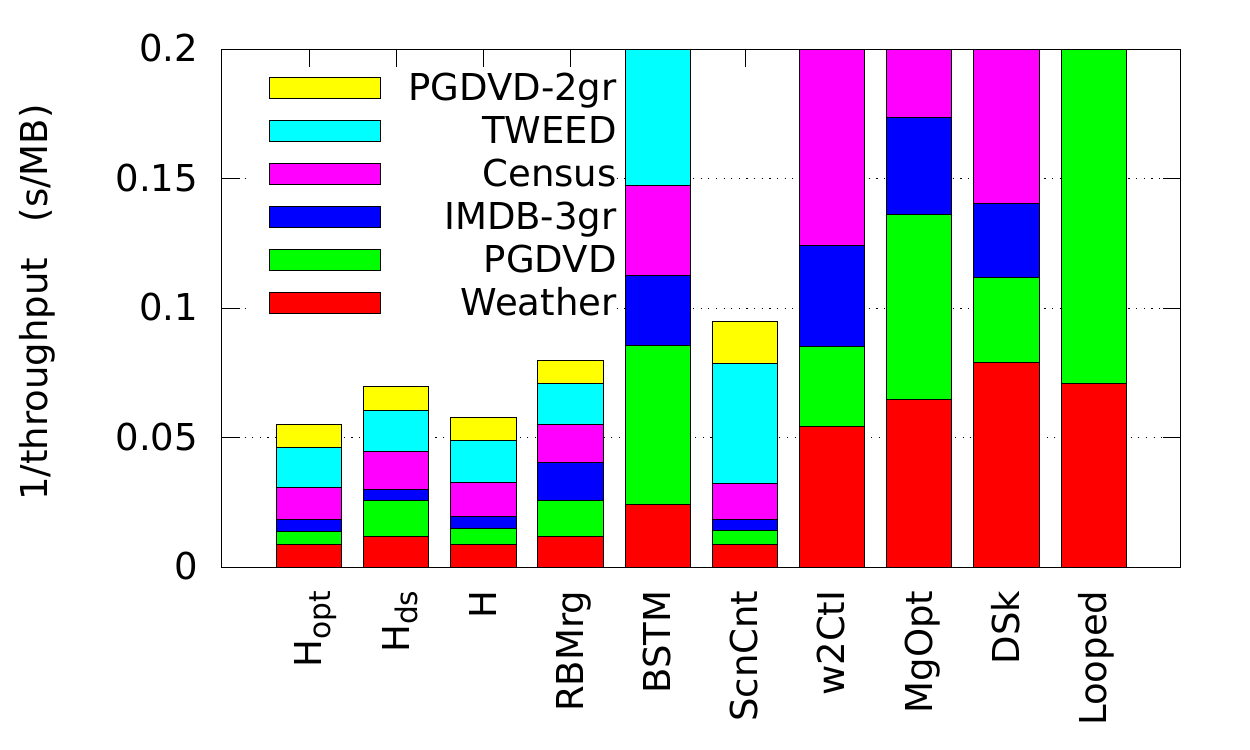}}
\subfloat [Text-derived datasets]{\includegraphics[width=.5\textwidth]{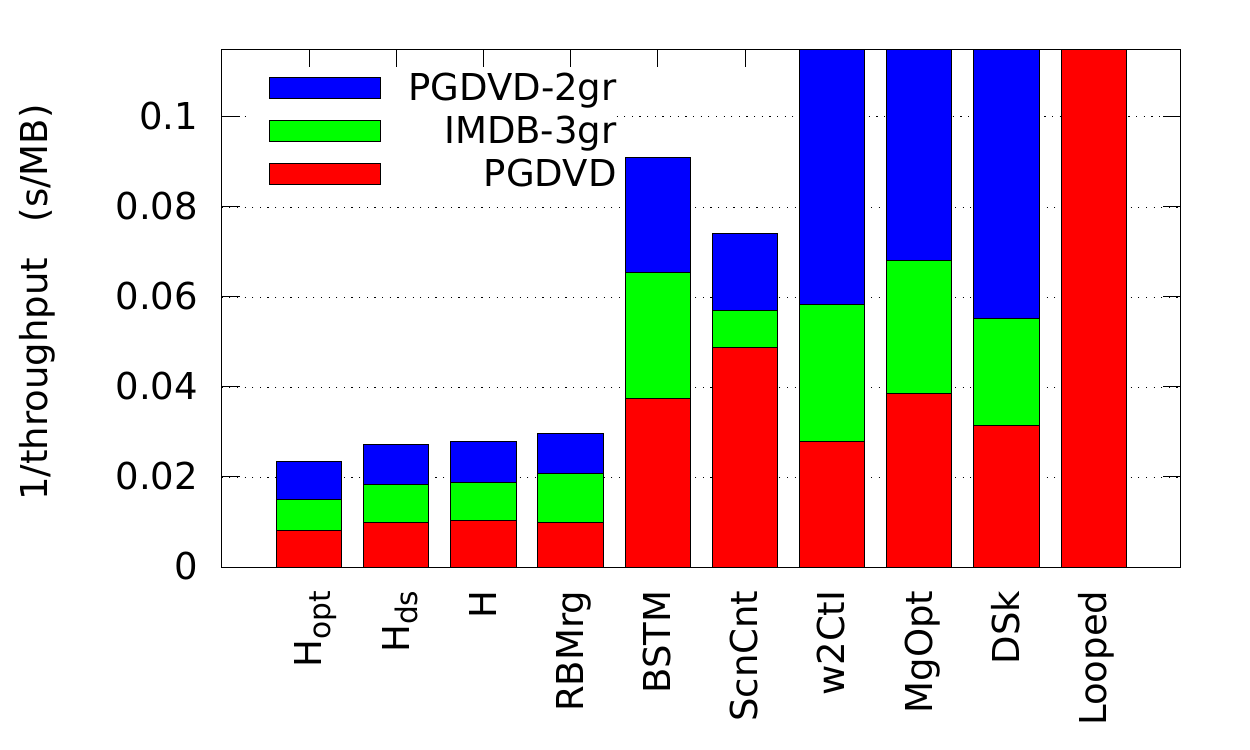}}
\caption{\label{fig:mixed-workload-barchart-hhelps}  Some cases when it is best to mix \cdom\ and \scncnt.
}
\end{figure}

Algorithm \cdom\ requires a detailed knowledge of the internal
workings of a RLE-compressed bitmap representation and is best added
by the maintainers of a compressed bitmap package.  Thus, it is
reasonable to look at the tradeoffs that come from 
hybrid algorithms that omit \cdom.
Our comparison
corresponds to bar heights in Fig.~\ref{fig:mixed-workload-barchart-hhelps}, and
the time for \h\ increased by \SI{66}{\percent} 
for the $N\geq 200$ case and \SI{154}{\percent} 
for the text-derived datasets.  Excluding \cdom, the best non-hybrid algorithm
was \scncnt.  When \h\ could not choose \cdom, its result was
\SI{2}{\percent} \emph{worse} than \scncnt\ on the text-derived datasets 
but \SI{6}{\percent} better for the $N\geq 200$ cases. 
Note that hybrid
algorithms can do much better: if \hopt\ chooses between \bstm, \scncnt\ and
\looped, on the text-derived datasets we can get
a result \SI{56}{\percent} better than \scncnt\  
(and only \SI{13}{\percent} worse than \cdom). 

\section{Conclusion and Future Work}

We reviewed several  novel and several known algorithms for computing thresholds. We found
that a novel algorithm (\cdom{}) was generally superior to  alternatives, 
sometimes being  orders of magnitude faster.

Although \cdom\ could be considered the overall
winner, each algorithm examined was weak in some circumstances.
However, 
we
combine them in a hybrid algorithm that improves on
any individual algorithm. 
In future work, we might create better hybrid algorithms,  perhaps by applying
machine-learning processes to choose the fastest threshold algorithm~\cite{lago:hybrid-algo-selection,horo:reasoning-resources}. 

Our work has considered $N$ values up to a few thousand (at most \num{11115}).  
Yet 
datasets whose indexes have millions of bitmaps are not
out of the question.  Would there be applications where a threshold
computation with $N=\num{1000000}$ would
be useful?  If so, which algorithms should be used?  
Can new algorithms be developed for this case?

When possible, data should be indexed in sorted order~\cite{arxiv:0901.3751}: this can improve compression and processing speed. 
Some algorithms might
benefit more than others from sorting, and this warrants further investigation.

Finally, algorithms can be parallelized, and while most of our
threshold computations take only a few milliseconds,  some
take tens of seconds.  If we try extremely large $N$
values, this may increase.  At some point, it may become important to
have one threshold computation run faster than is possible using a
single core.  For multicore processing, a particular challenge
with current architectures  is that
all cores compete for access to L3 and RAM\@. E.g., this means that it
is best if intermediate results fit in L2 cache.
It might be advisable to 
partition the problems.

\bibliographystyle{wileyj} 
\bibliography{symmetricbitmap}
\end{document}